\documentclass[10pt,english,allowdisplaybreaks,draftcls,onecolumn]{IEEEtran}
\usepackage[latin1]{inputenc}
\usepackage{geometry}
\usepackage{color}
\usepackage{amsthm}
\usepackage{amsmath}
\usepackage{amssymb}
\usepackage{graphicx}
\usepackage{setspace}
\usepackage{esint}
\makeatletter

\providecommand{\tabularnewline}{\\}

\theoremstyle{plain}
\newtheorem{thm}{\protect\theoremname}
\theoremstyle{plain}
\newtheorem{lem}[thm]{\protect\lemmaname}
\theoremstyle{plain}
\newtheorem{cor}[thm]{\protect\corollaryname}
\theoremstyle{plain}
\newtheorem{prop}[thm]{\protect\propositionname}

\@ifundefined{definecolor}
 {\usepackage{color}}{}
\@ifundefined{definecolor}{\usepackage{color}}{}

\usepackage{babel}

\usepackage{babel}

\usepackage{babel}
\providecommand{\theoremname}{Theorem}

\allowdisplaybreaks

\usepackage{babel}
\providecommand{\corollaryname}{Corollary}

\providecommand{\lemmaname}{Lemma}
\providecommand{\theoremname}{Theorem}

\usepackage{babel}
\providecommand{\corollaryname}{\inputencoding{latin9}Corollary}
\providecommand{\lemmaname}{\inputencoding{latin9}Lemma}
\providecommand{\propositionname}{\inputencoding{latin9}Proposition}
\providecommand{\theoremname}{\inputencoding{latin9}Theorem}

\makeatother

\usepackage{babel}
\providecommand{\corollaryname}{\inputencoding{latin9}Corollary}
\providecommand{\lemmaname}{\inputencoding{latin9}Lemma}
\providecommand{\propositionname}{\inputencoding{latin9}Proposition}
\providecommand{\theoremname}{\inputencoding{latin9}Theorem}

\begin{document}

\title{On the Achievable Rate-Regions for State-Dependent Gaussian Interference
Channel}

\author{\authorblockA{Shahab Ghasemi-Goojani and Hamid Behroozi\\
 Department of Electrical Engineering, Sharif University of Technology,
Tehran, Iran\\
 Email: shahab\_ghasemi@ee.sharif.edu, behroozi@sharif.edu}}
\maketitle
\begin{abstract}
In this paper, we study a general additive state-dependent Gaussian
interference channel (ASD-GIC) where we consider two-user interference
channel with two independent states known non-causally at both transmitters,
but unknown to either of the receivers. An special case, where the
additive states over the two links are the same is studied in \cite{Zhang_Allerton_2011,Zhang_IT_11},
in which it is shown that the gap between the achievable symmetric
rate and the upper bound is less than $\frac{1}{4}$ bit for the strong
interference case. Here, we also consider the case where interference
link gains satisfy $a_{12}\geq\frac{N_{1}}{N_{2}}$ and $a_{21}\geq\frac{N_{2}}{N_{1}}$
($N_{i}$ is the channel noise variance) and each channel state has
unbounded variance \cite{Philosof_11}, which is referred to as the
strong interferences. We first obtain an outer bound on the capacity
region. By utilizing lattice-based coding schemes, we obtain four
achievable rate regions. Depend on noise variance and channel power
constraint, achievable rate regions can coincide with the channel
capacity region. For the symmetric model, the achievable sum-rate
reaches to within 0.661 bit of the channel capacity for signal to
noise ratio (SNR) greater than one.
\end{abstract}

\section{Introduction}

An interference channel (IC) can be seen as a model for single-hop
multiple one-to-one communications, such as pairs of base stations
transmitting over a frequency band that suffers from intercell interference.
The earliest research on IC was initiated by Shannon \cite{Shannon_61}.
Unfortunately, the problem of characterizing the capacity region of
the general IC has been open for more than 30 years. Except for very
strong Gaussian IC, strong Gaussian IC and the sum-capacity for the
degraded Gaussian IC, characterizing the capacity region of a Gaussian
IC is still an open problem \cite{Sato_IT_81,Carleial_IT_75,Sason_IT_2004}.
By utilizing the superposition coding scheme, Carleial obtains an
achievable rate region \cite{Carleial_IT_78}. The best achievable
rate region known to date for a Gaussian IC, based on applying rate
splitting at the transmitters and simultaneous decoding at the receivers,
is established by Han and Kobayashi \cite{Han_IT_81}. Etkin et.al,
by deriving new outer bounds, show that an explicit Han-Kobayashi
version scheme can achieve capacity region within 1 bits for all channel
parameters \cite{Etkin_IT_2008}.

Many versions of the IC have also been studied in the literature,
including the IC with partial transmitter cooperation \cite{Maric_IT_2007},
the IC with conferencing encoders/decoders \cite{Cao_07_ISIT,Prabhakaran_IT_Cooperation_2011},
the Gaussian IC with feedback \cite{Changho_IT_2011} and the Gaussian
IC with potent relay \cite{Tian_IT_11}. In \cite{Zhang_Allerton_2011},
the two-user state-dependent symmetric Gaussian interference channel,
where the additive states over the two links are the same, is studied,
in which it is shown that the gap between the achievable symmetric
rate and the upper bound is less than $\frac{1}{4}$ bit for the strong
interference case and less than $\frac{3}{4}$ bit for the weak interference
case. In \cite{Zhang_WCNC_2011} an active interference cancellation
mechanism, which is a generalized dirty-paper coding technique, to
partially eliminate the effect of the state at the receivers is investigated.
It is shown that active interference cancellation significantly enlarges
the achievable rate-region.

In this paper, we study another type of the Gaussian IC: the state-dependent
two-user IC with two independent states known non-causally at both
transmitters, but unknown to either of the receivers. This situation
may arise in a multi-cell downlink communication scenario, where two
interested cells are interfering with each other and the mobiles suffer
from some independent interference (which can be from other neighboring
cells' base-stations and considered as an state) non-causally known
at each of the base-stations. In addition, we consider the interferences
as arbitrary, or equivalently Gaussians with unbounded variances,
and channel gains are larger than one (in the symmetric model), which
is refereed to as the strong interference \cite{Philosof_11}. We
provide an achievable rate-region based on lattice codes.

A comprehensive study on the performance of lattices is presented
in \cite{Zamir_09_ITA}. The problem of achieving an additive white
Gaussian noise (AWGN) channel capacity by utilizing lattice codes
is studied in \cite{Erez_IT_04}. A dirty paper AWGN channel in which
the interference is known non-causally or causally at the transmitter
is investigated in \cite{Ershi_IT_05}. For the non-causal case, it
is proved that the capacity loss due to applying the lattice strategy
for Gaussian noise is upper-bounded by $\frac{1}{2}\log\left(2\pi eG\left(\Lambda\right)\right)$,
where $G\left(\Lambda\right)$ is the normalized second moment of
the lattice. If the lattice code satisfies the following condition,$\underset{n\rightarrow\infty}{\lim}G\left(\Lambda\right)=\frac{1}{2\pi e}$,
this result coincides with the results of Costa's dirty-paper coding
(DPC) \cite{Costa_83}. In \cite{Philosof_11}, it is shown that the
lattice coding strategy may outperform the DPC in doubly dirty multiple-access
channel (MAC). By establishing an outer bound for doubly dirty MAC,
Wang is proved that the achievable rate-region by layered lattice
scheme is within a constant gap, which is independent of all channel
parameters, from the capacity region \cite{Wang_IT_2012}. In \cite{Ghasemi_12_PIMRC},
we also show that if the noise variance satisfy a constraint, then
the capacity region of an ASD-GIC with common channel state is achieved
when the state power goes to infinity.

In this work, we use a lattice-based coding scheme to obtain four
achievable rate regions for ASD-GIC. By comparing with an outer bound,
which is established for an asymptotic case, where the \textcolor{black}{channel
state }is assumed to be Gaussian with unbounded variance, we evaluate
each achievable rate region. We observe that the lattice based coding
scheme can achieve the capacity of the channel under some conditions,
which depend on the noise variance and power constraint of the channel.
For symmetric ASD-GIC, the achievable rate region of the lattice-based
scheme, dependent on the noise variance, is the capacity region or
is within 0.661 bit of the channel capacity for signal to noise ratio
(SNR) larger than one.

The remainder of the paper is organized as follows: We present the
channel model in Section \ref{sec:Preliminaries:-Lattices-and}. In
Section \ref{sec:Outer-bound}, an outer bound on the capacity region
is obtained. Lattice-based achievable rate-regions are presented in
Section \ref{sec:Lattice-Alignment}. Section \ref{sec:Conclusion}
concludes the paper.

\section{\label{sec:Preliminaries:-Lattices-and} Channel Model}

\subsection{Notations and Channel Model}

Throughout the paper, random variables and their realizations are
denoted by capital and small letters, respectively. $\boldsymbol{x}$
stands for a vector of length $n$, $(x_{1},x_{2},...,x_{n})$. Also,
$\left\Vert \boldsymbol{.}\right\Vert $ denotes the Euclidean norm,
and all logarithms are with respect to base $2$.

\begin{figure}
\begin{centering}
\includegraphics[width=12cm]{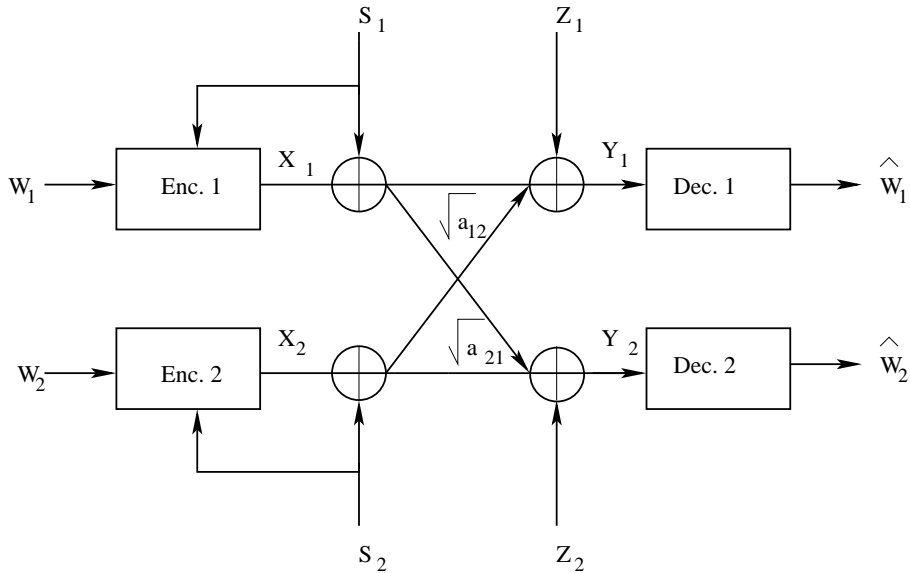} 
\par\end{centering}

\caption{\label{fig:The-Gaussian-interference}The Gaussian interference channel
with common interference known non-causally at both encoders.}
\end{figure}

In this paper, an additive state-dependent Gaussian interference channel
(ASD-GIC) where the channel states information are independent and
known non-causally at both encoders is considered. The system model
is depicted in Fig. \ref{fig:The-Gaussian-interference}. This channel
can be described by the following following equations (after suitable
normalization): 
\begin{eqnarray*}
\boldsymbol{Y}_{1} & = & \boldsymbol{X}_{1}+\sqrt{a_{12}}\boldsymbol{X}_{2}+\boldsymbol{S}_{1}+\sqrt{a_{12}}\boldsymbol{S}_{2}+\boldsymbol{Z}_{1},\\
\boldsymbol{Y}_{2} & = & \sqrt{a_{21}}\boldsymbol{X}_{1}+\boldsymbol{X}_{2}+\sqrt{a_{21}}\boldsymbol{S}_{1}+\boldsymbol{S}_{2}+\boldsymbol{Z}_{2},
\end{eqnarray*}
 where $\boldsymbol{X}_{i}$ is the channel input, $\boldsymbol{S}_{i}$
is an additive arbitrary distributed interference with variance $Q_{i}$
(or equivalently Gaussian with variance going to infinity), and $\boldsymbol{Z}_{i}$
represents an AWGN of mean zero and variance $N_{i}$. In this work,
we consider the strong Gaussian IC with state information, i.e., the
interference link gains satisfy $a_{12}\geq\frac{N_{1}}{N_{2}}$ and
$a_{21}\geq\frac{N_{2}}{N_{1}}$ \cite{Zhang_IT_11} and each channel
state has unbounded variance \cite{Philosof_11}.

The message $W_{i}$ at each encoder is mapped to $\boldsymbol{X}_{i}$
based on the non-causally known state information $\boldsymbol{S}_{i}$.
Note that $\left|\mathcal{W}_{1}\right|=2^{nR_{1}}$ and $\left|\mathcal{W}_{2}\right|=2^{nR_{2}}$.
Transmitted sequences $\boldsymbol{X}_{1}$, $\boldsymbol{X}_{2}$
are average-power limited to $P_{i}>0$, i.e., 
\begin{equation}
\frac{1}{n}\sum_{j=1}^{n}\mathbb{E}\left[\left|X_{i}[j]\right|^{2}\right]\leq P_{i},\,\,\,\,\,\textrm{f}\textrm{or}\,\,\, i=1,2.\label{PowerConstraint}
\end{equation}
 Each receiver needs to decode the information from the intended transmitter.
Based on the channel output, $\boldsymbol{Y}_{i}$, each receiver
makes an estimate of the corresponding message $W_{i}$ as $\hat{W}_{1}$.
The average error probability is defined as:

\[
P_{e}^{(n)}=\frac{1}{2^{n(R_{1}+R_{2})}}\underset{w_{1},w_{2}}{\sum}{\rm Pr}\left\{ \hat{W}_{1}\neq W_{1}\, or\,\hat{W}_{2}\neq W_{2}|(W_{1},W_{2})\,{\rm is}\,\,{\rm sent}\right\} ,
\]
 where $(W_{1},W_{2})$ is assumed to be uniformly distributed over
$\left\{ 1,2,...,2^{nR_{1}}\right\} \times\left\{ 1,2,...,2^{nR_{2}}\right\} $.
A rate pair $\left(R_{1},R_{2}\right)$ is achievable if there exist
a sequence of length-$n$ code $C^{n}\left(R_{1},R_{2}\right)$ such
that $P_{e}^{(n)}\rightarrow0$ as $n\rightarrow\infty$ \cite{Cover_book_2ndEdition}.

In the following, we present an outer bound over the capacity region
for $Q_{i}\rightarrow\infty$.

\section{\label{sec:Outer-bound}Outer bound}

To obtain an outer bound, we use the similar approach as \cite{Sato_IT_81}.
First, we assume that $\boldsymbol{S}_{2}$ is known at both decoders.
Thus, we can consider the following channel model:

\begin{eqnarray}
\boldsymbol{Y}_{1} & = & \boldsymbol{X}_{1}+\sqrt{a_{12}}\boldsymbol{X}_{2}+\boldsymbol{S}_{1}+\boldsymbol{Z}_{1},\label{eq:Reciever 1 with S_2 known}\\
\boldsymbol{Y}_{2} & = & \sqrt{a_{21}}\boldsymbol{X}_{1}+\boldsymbol{X}_{2}+\sqrt{a_{21}}\boldsymbol{S}_{1}+\boldsymbol{Z}_{2}.\nonumber 
\end{eqnarray}
 Now, by dividing $\boldsymbol{Y}_{2}$ over $\sqrt{a_{21}}$, we
get

\[
\boldsymbol{Y}_{2}^{'}\overset{\triangle}{=}\frac{\boldsymbol{Y}_{2}}{\sqrt{a_{21}}}=\boldsymbol{X}_{1}+\frac{\boldsymbol{X}_{2}}{\sqrt{a_{21}}}+\boldsymbol{S}_{1}+\frac{\boldsymbol{Z}_{2}}{\sqrt{a_{21}}}.
\]
 Using Fano's inequality, we know

\begin{equation}
h\left(W_{2}|\boldsymbol{Y}_{2}\right)\leq n\epsilon_{n},\label{eq:recovery M_2}
\end{equation}
 where $\epsilon_{n}\rightarrow0$ as $n\rightarrow\infty$. \textcolor{black}{By
using $W_{2}$ and $\boldsymbol{S}_{2}$, we can construct $\boldsymbol{X}_{2}$.
Similar to Sato's approach, we also construct noise $\boldsymbol{Z}_{2}^{'}\sim\mathcal{N}\left(0,N_{1}-\frac{N_{2}}{a_{21}}\right)$,
which is independent of $\boldsymbol{Z}_{1}$ and $\boldsymbol{Z}_{2}$.
Since $a_{21}\geq\frac{N_{2}}{N_{1}}$, we add $\left(\sqrt{a_{12}}-\frac{1}{\sqrt{a_{21}}}\right)\boldsymbol{X}_{2}$
and $\boldsymbol{Z}_{2}^{'}$ to $\boldsymbol{Y}_{2}^{'}$. Thus,
we have}

\begin{eqnarray}
\boldsymbol{Y}_{2}^{''} & \overset{\triangle}{=} & \boldsymbol{Y}_{2}^{'}+\left(\sqrt{a_{12}}-\frac{1}{\sqrt{a_{21}}}\right)\boldsymbol{X}_{2}+\boldsymbol{Z}_{2}^{'}\nonumber \\
 & = & \boldsymbol{X}_{1}+\sqrt{a_{12}}\boldsymbol{X}_{2}+\boldsymbol{S}_{1}+\boldsymbol{Z}_{2}^{'}+\frac{\boldsymbol{Z}_{2}}{\sqrt{a_{21}}}\nonumber \\
 & = & \boldsymbol{X}_{1}+\sqrt{a_{12}}\boldsymbol{X}_{2}+\boldsymbol{S}_{1}+\boldsymbol{Z}_{1}^{'},\label{eq:reform reciever 2-1}
\end{eqnarray}
 \textcolor{black}{where $\boldsymbol{Z}_{1}^{'}\sim\mathcal{N}\left(0,N_{1}\right)$.
}Therefore, by comparing (\ref{eq:Reciever 1 with S_2 known}) and
(\ref{eq:reform reciever 2-1}), we have

\begin{equation}
h\left(W_{1}|\boldsymbol{Y}_{2}\right)\leq n\epsilon_{n}.\label{eq:recovery M_1}
\end{equation}
 Thus, from (\ref{eq:recovery M_2}) and (\ref{eq:recovery M_1}),
we get

\begin{equation}
h\left(W_{1},W_{2}|\boldsymbol{Y}_{2}\right)\leq n\epsilon_{n}.\label{eq:recovery of both message}
\end{equation}
 Now, by considering the above-mentioned model, we obtain an outer
bound over the ASD-GIC capacity region. 
\begin{thm}
\label{thm:OuterRegion}In the limit of strong Gaussian interferences,
the capacity region of the ASD-GIC is contained in the following region:

\begin{equation}
R_{1}+R_{2}\leq\min\left(\frac{1}{2}\log\left(1+\frac{a_{12}P_{2}}{N_{1}}\right),\frac{1}{2}\log\left(1+\frac{a_{21}P_{1}}{N_{2}}\right)\right).\label{eq:Outer bound-1}
\end{equation}
 \end{thm}
\begin{IEEEproof}
We bound the sum rate $R_{1}+R_{2}$ as 
\begin{eqnarray}
n\left(R_{1}+R_{2}\right) & \leq & h\left(W_{1},W_{2}\right),\nonumber \\
 & = & h(W_{1},W_{2}|\boldsymbol{Y}_{2})+I(W_{1},W_{2};\boldsymbol{Y}_{2}),\nonumber \\
 & \leq & n\epsilon_{n}+I(W_{1},W_{2};\boldsymbol{Y}_{2}),\label{eq:Fano's Inequality}
\end{eqnarray}
 where (\ref{eq:Fano's Inequality}) follows from (\ref{eq:recovery of both message}).
Now, we assume that $\boldsymbol{S}_{2}$ is known at both decoders.
Then, we have 
\begin{eqnarray}
I\left(W_{1},W_{2};\boldsymbol{Y}_{2}\right) & = & h(\boldsymbol{Y}_{2})-h(\boldsymbol{Y}_{2}|W_{1},W_{2}),\nonumber \\
 & \leq & h(\boldsymbol{Y}_{2})-h(\boldsymbol{Y}_{2}|\boldsymbol{X}_{2},W_{1},W_{2}),\label{eq:Codition over entropy}\\
 & = & h(\boldsymbol{Y}_{2})-h(\boldsymbol{Y}_{2}|\boldsymbol{S}_{1},W_{1},W_{2},\boldsymbol{X}_{2})-I(\boldsymbol{S}_{1};\boldsymbol{Y}_{2}|W_{1},W_{2},\boldsymbol{X}_{2}),\nonumber \\
 & = & h(\boldsymbol{Y}_{2})-h(\boldsymbol{Z}_{2})-h(\boldsymbol{S}_{1}|W_{1},W_{2},\boldsymbol{X}_{2})+h(\boldsymbol{S}_{1}|W_{1},W_{2},\boldsymbol{X}_{2},\boldsymbol{Y}_{2}),\nonumber \\
 & \leq & h(\boldsymbol{Y}_{2})-h(\boldsymbol{Z}_{2})-h(\boldsymbol{S}_{1})+h(\boldsymbol{X}_{1}+\frac{\boldsymbol{Z}_{2}}{\sqrt{a_{21}}}),\label{eq:S_1 is Independent to (w_1,w_2,X_2}\\
 & \leq & \frac{n}{2}\log\left(\frac{N_{2}+\left(\sqrt{P_{2}}+\sqrt{a_{21}P_{1}}+\sqrt{a_{21}Q_{1}}\right)^{2}}{a_{21}Q_{1}}\right)+\frac{n}{2}\log\left(1+\frac{a_{21}P_{1}}{N_{2}}\right),\label{eq:Cauchy-Schwarz inequality}
\end{eqnarray}
 where (\ref{eq:Codition over entropy}) follows from the fact that
conditioning reduces entropy, (\ref{eq:S_1 is Independent to (w_1,w_2,X_2})
follows from the fact that $\boldsymbol{S}_{1}$ is independent to
$\left(W_{1},W_{2},\boldsymbol{X}_{2}\right)$ and (\ref{eq:Cauchy-Schwarz inequality})
follows from the fact that Gaussian distribution maximizes differential
entropy for a fixed second moment and Cauchy-Schwarz inequality. In
the limit of strong interference, i.e., $Q_{1}\rightarrow+\infty$,
we get 
\begin{equation}
R_{1}+R_{2}\leq\frac{1}{2}\log\left(1+\frac{a_{21}P_{1}}{N_{2}}\right).\label{eq:Outer bound 1-1}
\end{equation}
 Similarly, by assuming $\boldsymbol{S}_{1}$ is known at both decoders,
and reforming the above equations, we get

\begin{equation}
R_{1}+R_{2}\leq\frac{1}{2}\log\left(1+\frac{a_{12}P_{2}}{N_{1}}\right).\label{eq:Outer bound 2}
\end{equation}
 By combining (\ref{eq:Outer bound 1-1}) and (\ref{eq:Outer bound 2}),
we obtain

\begin{equation}
R_{1}+R_{2}\leq\min\left(\frac{1}{2}\log\left(1+\frac{a_{12}P_{2}}{N_{1}}\right),\frac{1}{2}\log\left(1+\frac{a_{21}P_{1}}{N_{2}}\right)\right).\label{eq:Outer bound}
\end{equation}

\end{IEEEproof}

\section{\label{sec:Achievable-Region-By}Achievable Rate-Regions}

\subsection{An Achievable Rate-Region based on Random Binning Scheme}

An achievable region for this channel can be obtained by random binning
argument. It is known that for the traditional strong Gaussian IC,
the capacity region is simply the intersection of two MAC rate-regions
\cite{Sato_IT_81}. Based on this idea, we can consider ASD-GIC as
two doubly-dirty MACs. The best rate-region for a doubly dirty MAC,
based on the random binning technique, is presented in \cite{Jafar_IT_06},
that is given by the convex hull of all rate pairs $(R_{1},R_{2})$
such that

\begin{eqnarray}
R_{1} & \leq & I\left(U_{1};Y|U_{2}\right)-I(U_{1};S_{1}),\nonumber \\
R_{2} & \leq & I\left(U_{2};Y|U_{1}\right)-I(U_{2};S_{2}),\nonumber \\
R_{1}+R_{2} & \leq & I\left(U_{1},U_{2};Y\right)-I(U_{1};S_{1})-I(U_{2};S_{2}),\label{eq:Sum rate In Jafar region}
\end{eqnarray}
 for some $p(u_{1},u_{2},x_{1},x_{2}|s_{1},s_{2})=p(u_{1},x_{1}|s_{1})p(u_{2},x_{2}|s_{2})$.
Now, for $Q_{i}\rightarrow\infty$, we can evaluate the achievable
sum rate in (\ref{eq:Sum rate In Jafar region}) for both MACs and
show that we cannot achieve any positive rates using such random binning
scheme. This evaluation is quite similar to the presented approach
in \cite{Philosof_11} and is provided in Appendix.

\subsection{\label{sec:Lattice-Alignment}Lattice Alignment}

\subsubsection{Lattice Definitions}

Here, we provide some necessary definitions on lattices and nested
lattice codes \cite{Erez_IT_04,Nazar_IT_11,conway_book}.

An $n$-\textit{\textcolor{black}{dimensional lattice}} $\Lambda$
is a set of points in Euclidean space $\mathbb{R}^{n}$ such that,
if $\boldsymbol{x},\boldsymbol{y}\in\Lambda$, then $\boldsymbol{x}+\boldsymbol{y}\in\Lambda$,
and if $\boldsymbol{x}\in\Lambda$ , then $-\boldsymbol{x}\in\Lambda$.
A lattice $\Lambda$ can always be written in terms of a generator
matrix $\mathbf{G}\in\mathbb{Z}^{n\times n}$ as 
\[
\Lambda=\{\boldsymbol{x}=\boldsymbol{z}\mathbf{G}:\boldsymbol{z}\in\mathbb{Z}^{n}\},
\]
 where $\mathbb{Z}$ represents integers.

The \textit{\textcolor{black}{nearest neighbor quantizer}} $\mathcal{Q}(.)$
associated with lattice $\Lambda$ is 
\[
\mathcal{Q}_{\Lambda}(\boldsymbol{x})=\arg\underset{\boldsymbol{l}\in\Lambda}{\min}\left\Vert \boldsymbol{x}-\boldsymbol{l}\right\Vert .
\]
 The \textit{\textcolor{black}{fundamental Voronoi region}} of lattice
$\Lambda$ is set of points in $\mathbb{R}^{n}$ closest to the zero
codeword, i.e.,

\[
\mathcal{V}_{0}(\Lambda)=\{\boldsymbol{x}\in\mathbb{\mathbb{R}}^{n}:\mathcal{Q}(\boldsymbol{x})=0\}.
\]
 $\sigma^{2}\left(\Lambda\right)$ which is called the second moment
of lattice $\Lambda$ is defined as 
\begin{equation}
\sigma^{2}(\Lambda)=\frac{1}{n}\frac{\int_{\mathcal{V}(\Lambda)}\left\Vert \boldsymbol{x}\right\Vert ^{2}d\boldsymbol{x}}{\int_{\mathcal{V}(\Lambda)}d\boldsymbol{x}},\label{eq:SM}
\end{equation}
 and the \textit{\textcolor{black}{normalized second moment}} of lattice
$\Lambda$ can be expressed as 
\[
G(\Lambda)=\frac{\sigma^{2}(\Lambda)}{[\int_{\mathcal{V}(\Lambda)}d\boldsymbol{x}]^{\frac{2}{n}}}=\frac{\sigma^{2}(\Lambda)}{V^{\frac{2}{n}}},
\]
 where $V=\int_{\mathcal{V}(\Lambda)}d\boldsymbol{x}$ is the Voronoi
region volume, i.e., $V=\mbox{ Vol}(\mathcal{V})$.

The \textit{\textcolor{black}{modulo-}}$\Lambda$ \textit{\textcolor{black}{operation}}
with respect to lattice $\Lambda$ is defined as 
\[
\boldsymbol{x}\mbox{ mod }\Lambda=\boldsymbol{x}-\mathcal{Q}(\boldsymbol{x}),
\]
 that maps $\boldsymbol{x}$ into a point in the fundamental Voronoi
region. The modulo lattice operation satisfi{}es the following distributive
property

\[
\left[\boldsymbol{x}\mbox{ mod }\Lambda+\boldsymbol{y}\right]\mbox{ mod }\Lambda=\left[\boldsymbol{x}+\boldsymbol{y}\right]\mbox{ mod }\Lambda.
\]
 A sequence of lattices $\Lambda^{(n)}\subseteq\mathbb{R}^{n}$ is
good for mean-squared error (MSE) quantization if 
\[
\underset{n\rightarrow\infty}{\lim}G\left(\Lambda^{(n)}\right)=\frac{1}{2\pi e}.
\]
 The sequence is indexed by the lattice dimension $n$. The existence
of such lattices is shown in \cite{Zamir_IT_96,Erez_IT_05}.

Let $\boldsymbol{Z}$ be a length-$n$$i.i.d$ Gaussian vector, $\boldsymbol{Z}\thicksim\mathcal{N}\left(\boldsymbol{0},\sigma_{Z}^{2}\boldsymbol{I}_{n}\right)$.
The volume-to-noise ratio of a lattice is given by 
\[
\mu\left(\Lambda,\epsilon\right)=\frac{\left(\mbox{ Vol}(\mathcal{V})\right)^{2/n}}{\sigma_{Z}^{2}},
\]
 where $\sigma_{Z}^{2}$ is chosen such that $\mbox{ Pr}\left\{ \boldsymbol{Z}\notin\mathcal{V}\right\} =\epsilon$
and $\boldsymbol{I}_{n}$ is an $n\times n$ identity matrix. A sequence
of lattices $\Lambda^{(n)}$ is Poltyrev-good if 
\[
\underset{n\rightarrow\infty}{\lim}\mu\left(\Lambda^{(n)},\epsilon\right)=2\pi e,\,\,\,\,\,\forall\epsilon\in\left(0,1\right)
\]
 and for fixed volume-to-noise ratio, greater than $2\pi e$, $\mbox{ Pr}\left\{ \boldsymbol{Z}\notin\mathcal{V}^{n}\right\} $
decays exponentially in $n$ . Poltyrev showed that a sequence of
such lattices exists \cite{Poltyrev_IT_94}. The existence of a sequence
of lattices $\Lambda^{(n)}$ which are good in both senses (i.e.,
simultaneously are Poltyrev-good and Rogers-good) is shown in \cite{Erez_IT_05}.

\textcolor{black}{We now calculate differential entropy of an $n$-dimensional
random vector $\boldsymbol{D}$ which is distributed uniformly over
fundamental Voronoi region. We have \cite{Zamir_IT_96}}

\textcolor{black}{{} 
\begin{eqnarray}
h\left(\boldsymbol{D}\right) & = & \log\left(V\right),\nonumber \\
 & = & \log\left(\frac{\sigma^{2}(\Lambda)}{G\left(\Lambda\right)}\right)^{n/2},\nonumber \\
 & \thickapprox & \frac{n}{2}\log\left(2\pi e\sigma^{2}(\Lambda)\right),\label{eq:EntropyDither}
\end{eqnarray}
 where the last approximation holds for lattices which are good for
quantization.}

In the following, we present a key property of dithered lattice codes. 
\begin{lem}
\textbf{The Crypto Lemma \cite{Forney_Allerton_2003,Erez_IT_04}}
Let $\mathbf{V}$ be a random vector with an arbitrary distribution
over $\mathbb{R}^{n}$. If $\boldsymbol{D}$ is independent of $\mathbf{V}$
and uniformly distributed over $\mathcal{V}$, then $(\mathbf{V}+\boldsymbol{D})\textrm{ mod }\Lambda$
is also independent of $\mathbf{V}$ and uniformly distributed over
$\mathcal{V}$. 
\begin{IEEEproof}
See Lemma 2 in \textbf{\cite{Forney_Allerton_2003}.} 
\end{IEEEproof}
\end{lem}

\subsubsection{Imbalanced ASD-GIC}
\begin{thm}
\label{thm:Imbalanced-SNRs:-Suppose}Imbalanced SNRs: Suppose that
$N_{1}\leq\sqrt{a_{12}P_{2}P_{1}}-a_{12}P_{2}$ and $N_{2}\leq\sqrt{a_{21}P_{2}P_{1}}-a_{21}P_{1}$.
The capacity region of an ASD-GIC in the limit of strong Gaussian
interferences, i.e., $\boldsymbol{S}_{i}\sim\mathcal{N}(0,Q_{i})$
and \textup{$Q_{i}\rightarrow+\infty$,} is given by: 
\begin{eqnarray*}
R_{1}+R_{2} & \leq & \min\left(\frac{1}{2}\log\left(1+\frac{a_{12}P_{2}}{N_{1}}\right),\frac{1}{2}\log\left(1+\frac{a_{21}P_{1}}{N_{2}}\right)\right).
\end{eqnarray*}
 \end{thm}
\begin{IEEEproof}
Based on the outer region in Section \ref{sec:Outer-bound}, the converse
is proved. Here, we show achievability of the following region using
a lattice-based coding scheme: 
\[
R_{1}+R_{2}\leq\frac{1}{2}\log\left(1+\frac{a_{12}P_{2}}{N_{1}}\right),
\]
 where $N_{1}\leq\sqrt{a_{12}P_{2}P_{1}}-a_{12}P_{2}$. Suppose that
there exist three lattices $\Lambda_{1}$, $\Lambda_{2}$ and $\Lambda_{3}=\sqrt{a_{12}}\Lambda_{2}$,
which are \textcolor{black}{good for quantization} ($\underset{n\rightarrow\infty}{\lim}G\left(\Lambda_{i}\right)=\frac{1}{2\pi e},\,\textrm{for\,\,\,}i=1,2,3\,)$,
such that 
\[
\sigma^{2}\left(\Lambda_{1}\right)=P_{1},\,\,\sigma^{2}\left(\Lambda_{2}\right)=P_{2},\,\,\,\textrm{and}\,\,\sigma^{2}\left(\Lambda_{3}\right)=a_{12}P_{2}.
\]
 User 1 and user 2 use lattices $\Lambda_{1}$ and $\Lambda_{2}$
with second moments $P_{1}$ and $P_{2}$, respectively. It is also
assumed that \textcolor{black}{$\boldsymbol{D}_{1}$} and \textcolor{black}{$\boldsymbol{D}_{2}$}
are t\textcolor{black}{wo independent dithers, where} \textcolor{black}{$\boldsymbol{D}_{1}$}
is uniformly distributed over the Voronoi region $\mathcal{V}_{1}$
and \textcolor{black}{$\boldsymbol{D}_{2}$ is uniformly distributed
over the Voronoi region} $\mathcal{V}_{2}$. Dithers are known at
the decoders.

First, we achieve the following corner point

\[
\left(R_{1},R_{2}\right)=\left(0,\frac{1}{2}\log\left(1+\frac{a_{12}P_{2}}{N_{1}}\right)\right),
\]
 where $a_{12}P_{2}\left(\frac{a_{12}P_{2}+N_{1}}{a_{12}P_{2}}\right)^{2}\leq P_{1}$.
We assume that $\Lambda_{3}=\alpha\Lambda_{1}$. The encoders send
\begin{eqnarray*}
\boldsymbol{X}_{1} & = & \left[-\boldsymbol{S}_{1}-\boldsymbol{D}_{1}\right]\textrm{ mod }\Lambda_{1},\\
\boldsymbol{X}_{2} & = & \left[\boldsymbol{V}_{2}-\alpha\boldsymbol{S}_{2}\right]\textrm{ mod }\Lambda_{2},
\end{eqnarray*}
 where $\boldsymbol{V}_{2}\sim\mbox{ Unif}\left(\mathcal{V}_{2}\right)$
carry the information for user 2, and dither \textcolor{black}{$\boldsymbol{D}_{1}$}
is known at the encoder of user 1. At the receiver of user 1, based
on the channel output, given by 
\[
\boldsymbol{Y}_{1}=\boldsymbol{X}_{1}+\sqrt{a_{12}}\boldsymbol{X}_{2}+\boldsymbol{S}_{1}+\sqrt{a_{12}}\boldsymbol{S}_{2}+\boldsymbol{Z}_{1}.
\]
 the following operation is performed: \textcolor{black}{{} 
\begin{eqnarray}
\boldsymbol{Y}_{d1} & = & \left[\alpha\boldsymbol{Y}_{1}+\alpha\boldsymbol{D}_{1}\right]\textrm{ mod }\Lambda_{3},\nonumber \\
 & = & \left[\alpha\left(\left[-\boldsymbol{S}_{1}-\boldsymbol{D}_{1}\right]\textrm{ mod }\Lambda_{1}+\sqrt{a_{12}}\boldsymbol{X}_{2}+\boldsymbol{S}_{1}+\sqrt{a_{12}}\boldsymbol{S}_{2}+\boldsymbol{Z}_{1}\right)+\alpha\boldsymbol{D}_{1}\right]\textrm{ mod }\Lambda_{3},\nonumber \\
 & = & \left[\sqrt{a_{12}}\boldsymbol{V}_{2}+\alpha\sqrt{a_{12}}\boldsymbol{X}_{2}-\sqrt{a_{12}}\left(\boldsymbol{V}_{2}-\alpha\boldsymbol{S}_{2}\right)+\alpha\boldsymbol{Z}_{1}-\alpha\mathcal{Q}_{\Lambda_{1}}(-\boldsymbol{S}_{1}-\boldsymbol{D}_{1})\right]\textrm{ mod }\Lambda_{3},\nonumber \\
 & = & \left[\sqrt{a_{12}}\boldsymbol{V}_{2}+\left(\alpha-1\right)\sqrt{a_{12}}\boldsymbol{X}_{2}+\alpha\boldsymbol{Z}_{1}\right]\textrm{ mod }\Lambda_{3},\label{eq::Lattice Alignment for R_2 case 1}\\
 & = & \left[\sqrt{a_{12}}\boldsymbol{V}_{2}+\boldsymbol{Z}_{eff}\right]\textrm{ mod }\Lambda_{3},\nonumber 
\end{eqnarray}
 where} 
\[
\boldsymbol{Z}_{eff}=\left[\left(\alpha-1\right)\sqrt{a_{12}}\boldsymbol{X}_{2}+\alpha\boldsymbol{Z}_{1}\right]\textrm{ mod }\Lambda_{3}.
\]
 (\ref{eq::Lattice Alignment for R_2 case 1}) follows from $\Lambda_{3}=\alpha\Lambda_{1}$,
thus we have $\alpha\mathcal{Q}_{\Lambda_{1}}(-\boldsymbol{S}_{1}-\boldsymbol{D}_{1}))\in\Lambda_{3}$,
i.e., the interference signal is aligned with $\Lambda_{3}$. Hence,
this term disappears after the modulo operation. To calculate rate
$R_{2}$, it is assumed that $\boldsymbol{V}_{2}\sim\mbox{ Unif}\left(\mathcal{V}_{2}\right)$.
We have 
\begin{eqnarray}
R_{2} & = & \frac{1}{n}I\left(\boldsymbol{V}_{2};\boldsymbol{Y}_{d1}\right),\nonumber \\
 & = & \frac{1}{n}\left\{ h(\boldsymbol{Y}_{d1})-h(\boldsymbol{Y}_{d1}|\boldsymbol{V}_{2})\right\} \nonumber \\
 & = & \frac{1}{2}\log\left(\frac{\sigma^{2}\left(\Lambda_{3}\right)}{G\left(\Lambda_{3}\right)}\right)-\frac{1}{n}h\left(\left[\left(\alpha-1\right)\sqrt{a_{12}}\boldsymbol{X}_{2}+\alpha\boldsymbol{Z}_{1}\right]\textrm{ mod }\Lambda_{3}\right),\label{eq:Crypto lemma over voronoi region (R_2)}\\
 & \geq & \frac{1}{2}\log\left(\frac{a_{12}P_{2}}{a_{12}\left(\alpha-1\right)^{2}P_{2}+\alpha^{2}N_{1}}\right)-\frac{1}{2}\log\left(2\pi eG\left(\Lambda_{3}\right)\right),\label{eq:Gaussian entropy}
\end{eqnarray}
 where (\ref{eq:Crypto lemma over voronoi region (R_2)}) follows
from the fact that $\sqrt{a_{12}}\boldsymbol{V}_{2}$ is uniform over
$\mathcal{V}_{3}=\sqrt{a_{12}}\mathcal{V}_{2}$; thus $\boldsymbol{Y}_{d1}$
is also uniformly distributed over $\mathcal{V}_{3}$ (crypto lemma)
and then we can apply (\ref{eq:EntropyDither}). (\ref{eq:Gaussian entropy})
follows from the fact that modulo operation reduces the second moment
and entropy is maximized by the normal distribution for a fixed second
moment.\textcolor{black}{{} Now, we need to find the coefficient
$\alpha$ such that minimizes the mean squared error (MSE) of the
effective noise $\boldsymbol{Z}_{{\rm eff}}$}. Hence, 
\[
\alpha_{{\rm MMSE}}=\frac{a_{12}P_{2}}{a_{12}P_{2}+N_{1}}.
\]
 Applying the optimal $\alpha$ and a good quantization lattice $\Lambda_{1}$,
we can achieve the following corner point 
\begin{equation}
(R_{1},R_{2})=\left(0,\frac{1}{2}\log\left(1+\frac{a_{12}P_{2}}{N_{1}}\right)\right).\label{eq:Corner point 1 for decoder 1}
\end{equation}
 Clearly, for $a_{12}P_{2}\left(\frac{a_{12}P_{2}+N_{1}}{a_{12}P_{2}}\right)^{2}=P_{1}$
the inner bound meets the outer bound (\ref{eq:Outer bound}). Also,
for $a_{12}P_{2}\left(\frac{a_{12}P_{2}+N_{1}}{a_{12}P_{2}}\right)^{2}\leq P_{1}$,
the outer bound remains $\frac{1}{2}\log\left(1+\frac{a_{12}P_{2}}{N_{1}}\right)$,
thus the outer bound is also achievable.

Now, we achieve the following corner point

\[
\left(R_{1},R_{2}\right)=\left(\frac{1}{2}\log\left(1+\frac{a_{12}P_{2}}{N_{1}}\right),0\right),
\]
 where $a_{12}P_{2}\left(\frac{a_{12}P_{2}+N_{1}}{a_{12}P_{2}}\right)^{2}\leq P_{1}$.
We assume that $\Lambda_{3}=\alpha\Lambda_{1}$. The encoders send
\begin{eqnarray*}
\boldsymbol{X}_{1} & = & \left[\boldsymbol{V}_{1}-\boldsymbol{S}_{1}\right]\textrm{ mod }\Lambda_{1},\\
\boldsymbol{X}_{2} & = & \left[-\alpha\boldsymbol{S}_{2}-\boldsymbol{D}_{2}\right]\textrm{ mod }\Lambda_{2}.
\end{eqnarray*}
 where $\boldsymbol{V}_{1}\sim\mbox{ Unif}\left(\mathcal{V}_{1}\right)$
carry the information for user 1, the dither \textcolor{black}{$\boldsymbol{D}_{2}$}
is known at the encoder of user 2. The signal at receiver 1 is given
by 
\[
\boldsymbol{Y}_{1}=\boldsymbol{X}_{1}+\sqrt{a_{12}}\boldsymbol{X}_{2}+\boldsymbol{S}_{1}+\sqrt{a_{12}}\boldsymbol{S}_{2}+\boldsymbol{Z}_{1}.
\]
 At the receiver, the following operation is performed: 
\begin{eqnarray}
\boldsymbol{Y}_{d1} & = & \left[\alpha\boldsymbol{Y}_{1}+\sqrt{a_{12}}\boldsymbol{D}_{2}\right]\textrm{ mod }\Lambda_{3},\nonumber \\
 & = & \left[\alpha\left(\left[\boldsymbol{V}_{1}-\boldsymbol{S}_{1}\right]\textrm{ mod }\Lambda_{1}+\sqrt{a_{12}}\boldsymbol{X}_{2}+\boldsymbol{S}_{1}+\sqrt{a_{12}}\boldsymbol{S}_{2}+\boldsymbol{Z}_{1}\right)+\sqrt{a_{12}}\boldsymbol{D}_{2}\right]\textrm{ mod }\Lambda_{3},\nonumber \\
 & = & \left[\alpha\boldsymbol{V}_{1}+\alpha\sqrt{a_{12}}\boldsymbol{X}_{2}-\sqrt{a_{12}}\left(-\alpha\boldsymbol{S}_{2}-\boldsymbol{D}_{2}\right)+\alpha\boldsymbol{Z}_{1}-\alpha\mathcal{Q}_{\Lambda_{1}}(\boldsymbol{V}_{1}-\boldsymbol{S}_{1})\right]\textrm{ mod }\Lambda_{3},\nonumber \\
 & = & \left[\alpha\boldsymbol{V}_{1}+\left(\alpha-1\right)\sqrt{a_{12}}\boldsymbol{X}_{2}+\alpha\boldsymbol{Z}_{1}\right]\textrm{ mod }\Lambda_{3},\label{eq:Lattice Alignment for R_2-Case2}\\
 & = & \left[\alpha\boldsymbol{V}_{1}+\boldsymbol{Z}_{eff}\right]\textrm{ mod }\Lambda_{3},\nonumber 
\end{eqnarray}
 where 
\[
Z_{eff}=\left[\left(\alpha-1\right)\sqrt{a_{12}}\boldsymbol{X}_{2}+\alpha\boldsymbol{Z}_{1}\right]\textrm{ mod }\Lambda_{3},
\]
 and (\ref{eq:Lattice Alignment for R_2-Case2}) is based on $\alpha\Lambda_{1}=\Lambda_{3}$.
To calculate the rate $R_{2}$, it is assumed that $\boldsymbol{V}_{1}\sim\mbox{ Unif}\left(\mathcal{V}_{1}\right)$.
We have 
\begin{eqnarray}
R_{1} & = & \frac{1}{n}I\left(\boldsymbol{V}_{1};\boldsymbol{Y}_{d1}\right),\nonumber \\
 & = & \frac{1}{n}\left\{ h(\boldsymbol{Y}_{d1})-h(\boldsymbol{Y}_{d1}|\boldsymbol{V}_{1})\right\} ,\nonumber \\
 & = & \frac{1}{2}\log\left(\frac{a_{12}P_{2}}{G\left(\Lambda_{3}\right)}\right)-\frac{1}{n}h\left(\left[\left(\alpha-1\right)\sqrt{a_{12}}\boldsymbol{X}_{2}+\alpha\boldsymbol{Z}_{1}\right]\textrm{ mod }\Lambda_{3}\right),\label{eq:Crypto lemma over voronoi region (R_2)-case 2}\\
 & \geq & \frac{1}{2}\log\left(\frac{a_{12}P_{2}}{\left(\alpha-1\right)^{2}a_{12}P_{2}+\alpha^{2}N_{1}}\right)-\frac{1}{2}\log\left(2\pi eG\left(\Lambda_{3}\right)\right),\label{eq:Iequality for maximum entropy-Case 2:R_2}
\end{eqnarray}
 where (\ref{eq:Crypto lemma over voronoi region (R_2)-case 2}) follows
from\textcolor{black}{{} $\alpha^{2}P_{1}=a_{12}P_{2}$} and the fact
that $\alpha\boldsymbol{V}_{1}$ is uniformly distributed over $\mathcal{V}_{3}$;
thus $\boldsymbol{Y}_{d1}$ is also uniform over $\mathcal{V}_{3}$
(crypto lemma). (\ref{eq:Iequality for maximum entropy-Case 2:R_2})
\textcolor{black}{is based on this fact that the second moment is
increased by removing modulo, and also for a fixed second moment,
Gaussian distribution maximizes differential entropy. By considering
MMSE coefficient that minimizes the MSE of the effective noise $\boldsymbol{Z}_{{\rm eff}}$,
i.e., $\alpha=\frac{a_{12}P_{2}}{a_{12}P_{2}+N_{1}}$ and applying
a good quantization lattice $\Lambda_{3}$, we can achieve the following
corner point} 
\begin{equation}
(R_{1},R_{2})=\left(\frac{1}{2}\log\left(1+\frac{a_{12}P_{2}}{N_{1}}\right),0\right).\label{eq:Corner point 2  for decoder 1}
\end{equation}
 Clearly, for $a_{12}P_{2}\left(\frac{a_{12}P_{2}+N_{1}}{a_{12}P_{2}}\right)^{2}=P_{1}$
the inner bound meets the outer bound (\ref{eq:Outer bound}). Likewise,
for $a_{12}P_{2}\left(\frac{a_{12}P_{2}+N_{1}}{a_{12}P_{2}}\right)^{2}\leq P_{1}$,
the outer bound remains $\frac{1}{2}\log\left(1+\frac{a_{12}P_{2}}{N_{1}}\right)$,
thus the outer bound is also achievable.

By using time sharing between two corner points, (\ref{eq:Corner point 1 for decoder 1})
and (\ref{eq:Corner point 2  for decoder 1}), for decoder 1, we can
achieve the following sum-rate region: 
\begin{equation}
R_{1}+R_{2}\leq\frac{1}{2}\log\left(1+\frac{a_{12}P_{2}}{N_{1}}\right).\label{eq:achievable rate for decoder 1}
\end{equation}
 If $N_{2}\leq\sqrt{a_{12}P_{2}P_{1}}-a_{12}P_{1}$, by similar analysis
at decoder 2, we have 
\begin{equation}
R_{1}+R_{2}\leq\frac{1}{2}\log\left(1+\frac{a_{21}P_{1}}{N_{2}}\right).\label{eq:achievable rate for decoder 2}
\end{equation}
 Therefore by using (\ref{eq:achievable rate for decoder 1}) and
(\ref{eq:achievable rate for decoder 2}), we get the following achievable
rate region for ASD-GIC: 
\begin{eqnarray*}
R_{1}+R_{2} & \leq & \min\left(\frac{1}{2}\log\left(1+\frac{a_{12}P_{2}}{N_{1}}\right),\frac{1}{2}\log\left(1+\frac{a_{21}P_{1}}{N_{2}}\right)\right).
\end{eqnarray*}

\end{IEEEproof}

\subsubsection{Nearly Balanced ASD-GIC}
\begin{thm}
\label{thm:If--and}If $N_{1}\geq\sqrt{a_{12}P_{2}P_{1}}-\min\left(a_{12}P_{2},P_{1}\right)$
and $N_{2}\geq\sqrt{a_{21}P_{2}P_{1}}-\min\left(a_{21}P_{1},P_{2}\right)$
for $P_{1}\neq a_{12}P_{2}$ and $a_{21}P_{1}\neq P_{2}$, then, the
following region is achievable for ASD-GIC:

\begin{equation}
R_{1}+R_{2}\leq\min\left(u.c.e\left\{ \left[\frac{1}{2}\log\left(\frac{P_{1}+a_{12}P_{2}+N_{1}}{2N_{1}+\left(\sqrt{P_{1}}-\sqrt{a_{12}}P_{2}\right)^{2}}\right)\right]^{+}\right\} ,u.c.e\left\{ \left[\frac{1}{2}\log\left(\frac{P_{2}+a_{21}P_{1}+N_{2}}{2N_{2}+\left(\sqrt{P_{2}}-\sqrt{a_{21}}P_{1}\right)^{2}}\right)\right]^{+}\right\} \right),\label{eq:Achievable rate region}
\end{equation}
 where the upper convex envelope is with respect to $P_{1}$ and $P_{2}$.\end{thm}
\begin{IEEEproof}
We use the lattice-based coding scheme. Suppose that there exist three
lattices $\Lambda_{1}$, $\Lambda_{2}$ and $\Lambda_{3}=\sqrt{a_{12}}\Lambda_{2}$,
which are \textcolor{black}{good for quantization} ($\underset{n\rightarrow\infty}{\lim}G\left(\Lambda_{i}\right)=\frac{1}{2\pi e},\,\textrm{for\,\,\,}i=1,2,3\,)$,
such that 
\[
\sigma^{2}\left(\Lambda_{1}\right)=P_{1},\,\,\sigma^{2}\left(\Lambda_{2}\right)=P_{2},\,\,\,\textrm{and}\,\,\sigma^{2}\left(\Lambda_{3}\right)=a_{12}P_{2}.
\]
 User 1 and user 2 use the lattices $\Lambda_{1}$ and $\Lambda_{2}$
with second moments $P_{1}$ and $P_{2}$, respectively. It is also
assumed that \textcolor{black}{$\boldsymbol{D}_{1}$} and \textcolor{black}{$\boldsymbol{D}_{2}$}
are t\textcolor{black}{wo independent dithers} that \textcolor{black}{$\boldsymbol{D}_{1}$}
is uniformly distributed over the Voronoi region $\mathcal{V}_{1}$
and \textcolor{black}{$\boldsymbol{D}_{2}$ is uniformly distributed
over the Voronoi region} $\mathcal{V}_{2}$. Dithers are known at
the decoders.

First, we consider $a_{12}P_{2}\leq\frac{\left(P_{1}+N_{1}\right)^{2}}{P_{1}}$
or equivalently $N_{1}\geq\sqrt{a_{12}P_{1}P_{2}}-P_{1}$. We assume
that $\Lambda_{3}=\frac{\alpha_{2}}{\alpha_{1}}\Lambda_{1}$. The
encoders send 
\begin{eqnarray*}
\boldsymbol{X}_{1} & = & \left[-\alpha_{1}\boldsymbol{S}_{1}+\boldsymbol{D}_{1}\right]\textrm{ mod }\Lambda_{1},\\
\boldsymbol{X}_{2} & = & \left[\boldsymbol{V}_{2}-\alpha_{2}\boldsymbol{S}_{2}-\boldsymbol{D}_{2}\right]\textrm{ mod }\Lambda_{2}.
\end{eqnarray*}
 At the receiver of user 1, based on the channel output given by 
\[
\boldsymbol{Y}_{1}=\boldsymbol{X}_{1}+\sqrt{a_{12}}\boldsymbol{X}_{2}+\boldsymbol{S}_{1}+\sqrt{a_{12}}\boldsymbol{S}_{2}+\boldsymbol{Z}_{1},
\]
 the following operation is performed: 
\begin{eqnarray}
\boldsymbol{Y}_{d1} & = & \left[\alpha_{2}\boldsymbol{Y}_{1}+\sqrt{a_{12}}\boldsymbol{D}_{2}-\frac{\alpha_{2}}{\alpha_{1}}\boldsymbol{D}_{1}\right]\textrm{ mod }\Lambda_{3},\nonumber \\
 & = & \left[\alpha_{2}\left(\left[-\alpha_{1}\boldsymbol{S}_{1}+\boldsymbol{D}_{1}\right]\textrm{ mod }\Lambda_{1}+\sqrt{a_{12}}\boldsymbol{X}_{2}+\boldsymbol{S}_{1}+\sqrt{a_{12}}\boldsymbol{S}_{2}+\boldsymbol{Z}_{1}\right)+\sqrt{a_{12}}\boldsymbol{D}_{2}-\frac{\alpha_{2}}{\alpha_{1}}\boldsymbol{D}_{1}\right]\textrm{ mod }\Lambda_{3},\nonumber \\
 & = & \left[\sqrt{a_{12}}\boldsymbol{V}_{2}+\alpha_{2}\left(\sqrt{a_{12}}\boldsymbol{X}_{2}+\boldsymbol{Z}_{1}\right)-\sqrt{a_{12}}\left(\boldsymbol{V}_{2}-\alpha_{2}\boldsymbol{S}_{2}-\boldsymbol{D}_{2}\right)-\left(1-\alpha_{1}\right)\frac{\alpha_{2}}{\alpha_{1}}\left(-\alpha_{1}\boldsymbol{S}_{1}+\boldsymbol{D}_{1}\right)\right.\nonumber \\
 &  & \left.-\alpha_{2}\mathcal{Q}_{\Lambda_{1}}\left(-\alpha_{1}\boldsymbol{S}_{1}+\boldsymbol{D}_{1}\right)\right]\textrm{ mod }\Lambda_{3},\nonumber \\
 & = & \left[\sqrt{a_{12}}\boldsymbol{V}_{2}+\sqrt{a_{12}}\left(\alpha_{2}-1\right)\boldsymbol{X}_{2}-\left(1-\alpha_{1}\right)\frac{\alpha_{2}}{\alpha_{1}}\boldsymbol{X}_{1}+\alpha_{2}\boldsymbol{Z}_{1}-\frac{\alpha_{2}}{\alpha_{1}}\mathcal{Q}_{\Lambda_{1}}\left(-\alpha_{1}\boldsymbol{S}_{1}+\boldsymbol{D}_{1}\right)\right]\textrm{ mod }\Lambda_{3},\label{eq:Distributive law, R_2, Case 1}\\
 & = & \left[\sqrt{a_{12}}\boldsymbol{V}_{2}+\sqrt{a_{12}}\left(\alpha_{2}-1\right)\boldsymbol{X}_{2}-\left(1-\alpha_{1}\right)\frac{\alpha_{2}}{\alpha_{1}}\boldsymbol{X}_{1}+\alpha_{2}\boldsymbol{Z}_{1}\right]\textrm{ mod }\Lambda_{3},\label{eq:Lattice Alignment for case 1:R_2}\\
 & = & \left[\sqrt{a_{12}}\boldsymbol{V}_{2}+\boldsymbol{Z}_{eff}\right]\textrm{ mod }\Lambda_{3},\nonumber 
\end{eqnarray}
 where 
\[
\boldsymbol{Z}_{eff}=\left[\sqrt{a_{12}}\left(\alpha_{2}-1\right)\boldsymbol{X}_{2}-\left(1-\alpha_{1}\right)\frac{\alpha_{2}}{\alpha_{1}}\boldsymbol{X}_{1}+\alpha_{2}\boldsymbol{Z}_{1}\right]\textrm{ mod }\Lambda_{3}.
\]
 (\ref{eq:Distributive law, R_2, Case 1}) is based on distributive
law and (\ref{eq:Lattice Alignment for case 1:R_2}) follows from
$\Lambda_{3}=\frac{\alpha_{2}}{\alpha_{1}}\Lambda_{1}$, we have that
$\frac{\alpha_{2}}{\alpha_{1}}\mathcal{Q}_{\Lambda_{1}}(-\alpha_{1}\boldsymbol{S}_{1}+\boldsymbol{D}_{1})\in\Lambda_{3}$,
i.e., the interference signal is aligned with $\Lambda_{3}$. Hence,
the element disappears after the modulo operation. To calculate the
rate $R_{2}$, it is assumed that $\boldsymbol{V}_{2}\sim\mbox{ Unif}\left(\mathcal{V}_{2}\right)$.
We have 
\begin{eqnarray}
R_{2} & = & \frac{1}{n}I\left(\boldsymbol{V}_{2};\boldsymbol{Y}_{d1}\right),\nonumber \\
 & = & \frac{1}{n}\left\{ h(\boldsymbol{Y}_{d1})-h(\boldsymbol{Y}_{d1}|\boldsymbol{V}_{2})\right\} \nonumber \\
 & = & \frac{1}{2}\log\left(\frac{a_{12}P_{2}}{G\left(\Lambda_{3}\right)}\right)-\frac{1}{n}h\left(\left[\sqrt{a_{12}}\left(\alpha_{2}-1\right)\boldsymbol{X}_{2}-\left(1-\alpha_{1}\right)\frac{\alpha_{2}}{\alpha_{1}}\boldsymbol{X}_{1}+\alpha_{2}\boldsymbol{Z}_{1}\right]\textrm{ mod }\Lambda_{3}\right),\label{eq:Crypto lemma for case 1:R_2}\\
 & \geq & \frac{1}{2}\log\left(\frac{a_{12}P_{2}}{a_{12}\left(\alpha_{2}-1\right)^{2}P_{2}+\left(\left(1-\alpha_{1}\right)\frac{\alpha_{2}}{\alpha_{1}}\right)^{2}P_{1}+\alpha_{2}^{2}N_{1}}\right)-\frac{1}{2}\log\left(2\pi eG\left(\Lambda_{3}\right)\right),\label{eq:Maximum entropy, Case1:R_2}
\end{eqnarray}
 where (\ref{eq:Crypto lemma for case 1:R_2}) follows from $\sqrt{a_{12}}\boldsymbol{V}_{2}$
is uniform over $\sqrt{a_{12}}\mathcal{V}_{2}$ therefore $\boldsymbol{Y}_{d1}$
is also uniform over $\sqrt{a_{12}}\mathcal{V}_{2}$ (crypto lemma)
and (\ref{eq:Maximum entropy, Case1:R_2}) follows from the fact that
modulo operation reduces the second moment an\textcolor{black}{d for
a fixed second moment, Gaussian distribution maximizes differential
entropy. }Now, by considering $\left(\frac{\alpha_{2}}{\alpha_{1}}\right)^{2}P_{1}=a_{12}P_{2}$,
and MMSE coefficient $\alpha_{2}$, such that the MSE of the effective
noise $\boldsymbol{Z}_{{\rm eff}}$ is minimized when the lattice
dimension goes to infinity, we obtain 
\[
\alpha_{{\rm 2,MMSE}}=\frac{\sqrt{a_{12}P_{2}}\left(\sqrt{P_{1}}+\sqrt{a_{12}P_{2}}\right)}{P_{1}+a_{12}P_{2}+N_{1}}.
\]
 With this chosen for $\alpha_{2}$, we get that the following rate
is achievable:

\begin{equation}
R_{2}\leq\left[\frac{1}{2}\log\left(\frac{P_{1}+a_{12}P_{2}+N_{1}}{2N_{1}+\left(\sqrt{P_{1}}-\sqrt{a_{12}P_{2}}\right)^{2}}\right)\right]^{+}.\label{eq:Case 1, Achievable rate 1}
\end{equation}
 Now, we consider $P_{1}\leq\frac{\left(a_{12}P_{2}+N_{1}\right)^{2}}{a_{12}P_{2}}$
or equivalently $N_{1}\geq\sqrt{a_{12}P_{2}P_{1}}-a_{12}P_{2}$. We
assume that $\Lambda_{3}=\frac{\alpha_{2}}{\alpha_{1}}\Lambda_{1}$.
The encoders send 
\begin{eqnarray*}
\boldsymbol{X}_{1} & = & \left[-\alpha_{1}\boldsymbol{S}_{1}+\boldsymbol{D}_{1}\right]\textrm{ mod }\Lambda_{1},\\
\boldsymbol{X}_{2} & = & \left[\boldsymbol{V}_{2}-\alpha_{2}\boldsymbol{S}_{2}+\boldsymbol{D}_{2}\right]\textrm{ mod }\Lambda_{2}.
\end{eqnarray*}
 At the receiver of user 1, based on the channel output given by 
\[
\boldsymbol{Y}_{1}=\boldsymbol{X}_{1}+\sqrt{a_{12}}\boldsymbol{X}_{2}+\boldsymbol{S}_{1}+\sqrt{a_{12}}\boldsymbol{S}_{2}+\boldsymbol{Z}_{1}.
\]
 The following operation is performed: 
\begin{eqnarray}
\boldsymbol{Y}_{d1} & = & \left[\alpha_{1}\boldsymbol{Y}_{1}-\frac{\alpha_{1}}{\alpha_{2}}\sqrt{a_{12}}\boldsymbol{D}_{2}-\boldsymbol{D}_{1}\right]\textrm{ mod }\Lambda_{1},\nonumber \\
 & = & \left[\alpha_{1}\left(\boldsymbol{X}_{1}+\sqrt{a_{12}}\boldsymbol{X}_{2}+\boldsymbol{S}_{1}+\sqrt{a_{12}}\boldsymbol{S}_{2}+\boldsymbol{Z}_{1}\right)-\frac{\alpha_{1}}{\alpha_{2}}\sqrt{a_{12}}\boldsymbol{D}_{2}-\boldsymbol{D}_{1}\right]\textrm{ mod }\Lambda_{1},\nonumber \\
 & = & \left[\frac{\alpha_{1}}{\alpha_{2}}\sqrt{a_{12}}\boldsymbol{V}_{2}+\alpha_{1}\left(\boldsymbol{X}_{1}+\sqrt{a_{12}}\boldsymbol{X}_{2}+\boldsymbol{Z}_{1}\right)-\left(-\alpha_{1}\boldsymbol{S}_{1}+\boldsymbol{D}_{1}\right)-\frac{\alpha_{1}}{\alpha_{2}}\sqrt{a_{12}}\left(\boldsymbol{V}_{2}-\alpha_{2}\boldsymbol{S}_{2}+\boldsymbol{D}_{2}\right)\right]\textrm{ mod }\Lambda_{1},\nonumber \\
 & = & \left[\frac{\alpha_{1}}{\alpha_{2}}\sqrt{a_{12}}\boldsymbol{V}_{2}+\left(\alpha_{1}-1\right)\boldsymbol{X}_{1}-\left(1-\alpha_{2}\right)\frac{\alpha_{1}}{\alpha_{2}}\sqrt{a_{12}}\boldsymbol{X}_{2}+\alpha_{1}\boldsymbol{Z}_{1}\right]\textrm{ mod }\Lambda_{1},\label{eq:Distributive law for case 2:R_2}\\
 & = & \left[\frac{\alpha_{1}}{\alpha_{2}}\sqrt{a_{12}}\boldsymbol{V}_{2}+\boldsymbol{Z}_{eff}\right]\textrm{ mod }\Lambda_{1},\nonumber 
\end{eqnarray}
 where 
\[
\boldsymbol{Z}_{eff}=\left[\left(\alpha_{1}-1\right)\boldsymbol{X}_{1}-\left(1-\alpha_{2}\right)\frac{\alpha_{1}}{\alpha_{2}}\sqrt{a_{12}}\boldsymbol{X}_{2}+\alpha_{1}\boldsymbol{Z}_{1}\right]\textrm{ mod }\Lambda_{1}.
\]
 (\ref{eq:Distributive law for case 2:R_2}) is based on distributive
law and follows from $\Lambda_{1}=\frac{\alpha_{1}}{\alpha_{2}}\Lambda_{3}$,
we have that $\frac{\alpha_{1}}{\alpha_{2}}\mathcal{Q}_{\Lambda_{3}}(\boldsymbol{V}_{2}-\alpha_{2}\boldsymbol{S}_{1}+\boldsymbol{D}_{2})\in\Lambda_{1}$.
Hence, the element disappears after the modulo operation. To calculate
the rate $R_{2}$, it is assumed that $\boldsymbol{V}_{2}\sim Unif\left(\mathcal{V}_{2}\right)$.
We have 
\begin{eqnarray}
R_{2} & = & \frac{1}{n}I\left(\boldsymbol{V}_{2};\boldsymbol{Y}_{d1}\right),\nonumber \\
 & = & \frac{1}{n}\left\{ h(\boldsymbol{Y}_{d1})-h(\boldsymbol{Y}_{d1}|\boldsymbol{V}_{2})\right\} \nonumber \\
 & = & \frac{1}{2}\log\left(\frac{P_{1}}{G\left(\Lambda_{1}\right)}\right)-\frac{1}{n}h\left(\left[\left(\alpha_{1}-1\right)\boldsymbol{X}_{1}-\left(1-\alpha_{2}\right)\frac{\alpha_{1}}{\alpha_{2}}\sqrt{a_{12}}\boldsymbol{X}_{2}+\alpha_{1}\boldsymbol{Z}_{1}\right]\textrm{ mod }\Lambda_{1}\right),\label{eq:Crypto lemma for case 2:R_2}\\
 & \geq & \frac{1}{2}\log\left(\frac{P_{1}}{\left(\alpha_{1}-1\right)^{2}P_{1}+\left(\left(1-\alpha_{2}\right)\frac{\alpha_{1}}{\alpha_{2}}\right)^{2}a_{12}P_{2}+\alpha_{1}^{2}N_{1}}\right)-\frac{1}{2}\log\left(2\pi eG\left(\Lambda_{1}\right)\right),\label{eq:Maximum entropy for case 2:R_2}
\end{eqnarray}
 where (\ref{eq:Distributive law for case 2:R_2}) follows from the
fact that $\sqrt{a_{12}}\boldsymbol{V}_{2}$ is uniform over $\mathcal{V}_{1}=\sqrt{a_{12}}\mathcal{V}_{2}$;
thus, $\boldsymbol{Y}_{d1}$ is also uniform over $\mathcal{V}_{1}$
(crypto lemma). Since modulo operation reduces the second moment a\textcolor{black}{nd
for a fixed second moment, the entropy is maximized by Gaussian distribution,}
(\ref{eq:Maximum entropy for case 2:R_2}) is correct\textcolor{black}{.}
Now, by considering $\left(\frac{\alpha_{1}}{\alpha_{2}}\right)^{2}a_{12}P_{2}=P_{1}$,
we find the optimal $\alpha$ when the lattice dimension goes to infinity
such that minimizes the MSE of the effective noise $\boldsymbol{Z}_{{\rm eff}}$
. Hence, 
\[
\alpha_{{\rm 1,MMSE}}=\frac{\sqrt{P_{1}}\left(\sqrt{P_{1}}+\sqrt{a_{12}P_{2}}\right)}{P_{1}+a_{12}P_{2}+N_{1}}.
\]
 With this chosen for $\alpha$, we get that any rate

\begin{equation}
R_{2}\leq\left[\frac{1}{2}\log\left(\frac{P_{1}+a_{12}P_{2}+N_{1}}{2N_{1}+\left(\sqrt{P_{1}}-\sqrt{a_{12}P_{2}}\right)^{2}}\right)\right]^{+},\label{eq:Case 2, Achievable rate 1}
\end{equation}
 is achievable. From (\ref{eq:Case 1, Achievable rate 1}) and (\ref{eq:Case 2, Achievable rate 1}),
we get the following corner point is achievable

\begin{equation}
\left(R_{1},R_{2}\right)=\left(0,\left[\frac{1}{2}\log\left(\frac{P_{1}+a_{12}P_{2}+N_{1}}{2N_{1}+\left(\sqrt{P_{1}}-\sqrt{a_{12}P_{2}}\right)^{2}}\right)\right]^{+}\right),\label{eq:Decoder 1, Corner point 1}
\end{equation}
 if

\[
N_{1}\geq\sqrt{a_{12}P_{1}P_{2}}-\min\left(a_{12}P_{2},P_{1}\right).
\]
 By symmetry, it can be shown that for $N_{1}\geq\sqrt{a_{12}P_{1}P_{2}}-\min\left(a_{12}P_{2},P_{1}\right)$,
the following corner point is achievable (see Appendix 2):

\begin{equation}
\left(R_{1},R_{2}\right)=\left(\left[\frac{1}{2}\log\left(\frac{P_{1}+a_{12}P_{2}+N_{1}}{2N_{1}+\left(\sqrt{P_{1}}-\sqrt{a_{12}P_{2}}\right)^{2}}\right)\right]^{+},0\right).\label{eq:Decoder 1. Corner point 2}
\end{equation}
 Thus, for decoder 1, we can achieve the following region by time
sharing between two corner points, (\ref{eq:Decoder 1, Corner point 1})
and (\ref{eq:Decoder 1. Corner point 2}):

\begin{equation}
R_{1}+R_{2}\leq\left[\frac{1}{2}\log\left(\frac{P_{1}+a_{12}P_{2}+N_{1}}{2N_{1}+\left(\sqrt{P_{1}}-\sqrt{a_{12}P_{2}}\right)^{2}}\right)\right]^{+},\label{eq:Achievable region for Decoder 1}
\end{equation}
 if

\[
N_{1}\geq\sqrt{a_{12}P_{1}P_{2}}-\min\left(a_{12}P_{2},P_{1}\right).
\]
 By similar analysis for decoder 2, we can achieve the following region

\begin{equation}
R_{1}+R_{2}\leq\left[\frac{1}{2}\log\left(\frac{a_{21}P_{1}+P_{2}+N_{2}}{2N_{2}+\left(\sqrt{P_{2}}-\sqrt{a_{21}P_{1}}\right)^{2}}\right)\right]^{+},\label{eq:Achievable region for decoder 2}
\end{equation}
 if

\[
N_{2}\geq\sqrt{a_{21}P_{1}P_{2}}-\min\left(a_{21}P_{1},P_{2}\right).
\]
 The theorem follows from (\ref{eq:Achievable region for Decoder 1})
and (\ref{eq:Achievable region for decoder 2}). 
\end{IEEEproof}

\subsubsection{Calculating the gap}

\begin{figure}
\begin{centering}
\includegraphics[width=12cm]{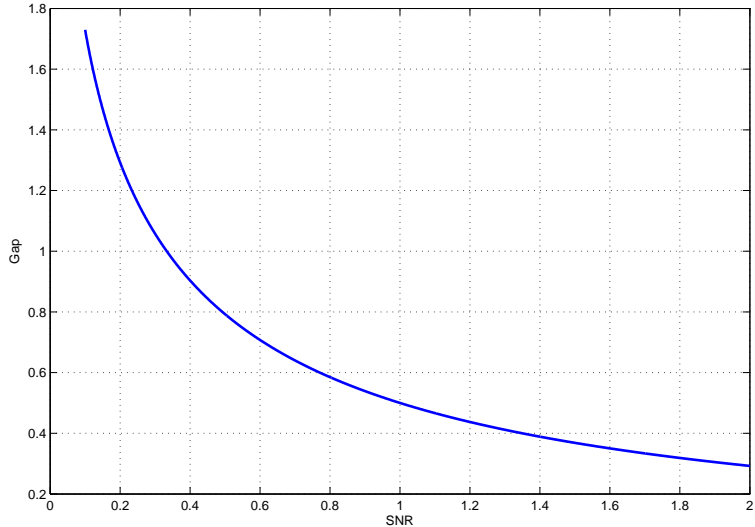} 
\par\end{centering}

\centering{}\caption{\label{fig:Gap-is-given}The gap {\small between the outer bound and
the achievable rate}, given by (\ref{eq:Gap between inner bound and outer bound}).}
\end{figure}

Now, we obtain the gap between the outer bound in (\ref{eq:Outer bound})
and the achievable rate region, given by (\ref{eq:Achievable rate region}).
First, we define the following gap:

\begin{eqnarray}
\xi\left(P_{1},P_{2},N_{1},N_{2},a_{12},a_{21}\right) & = & \min\left(\frac{1}{2}\log\left(1+\frac{a_{12}P_{2}}{N_{1}}\right),\frac{1}{2}\log\left(1+\frac{a_{21}P_{1}}{N_{2}}\right)\right)\nonumber \\
 & - & \min\left(u.c.e\left\{ \left[\frac{1}{2}\log\left(\frac{P_{1}+a_{12}P_{2}+N_{1}}{2N_{1}+\left(\sqrt{P_{1}}-\sqrt{a_{12}P_{2}}\right)^{2}}\right)\right]^{+}\right\} \right.\nonumber \\
 &  & \left.u.c.e\left\{ \left[\frac{1}{2}\log\left(\frac{P_{2}+a_{21}P_{1}+N_{2}}{2N_{2}+\left(\sqrt{P_{2}}-\sqrt{a_{21}P_{1}}\right)^{2}}\right)\right]^{+}\right\} \right).\label{eq:Gap between outer bound and achievable rate}
\end{eqnarray}
 Since, it is difficult to calculate the gap (\ref{eq:Gap between outer bound and achievable rate})
for asymmetric model, thus, we focus over symmetric model, i.e., $P_{1}=P_{2}\overset{\triangle}{=}P$,
$N_{1}=N_{2}\overset{\triangle}{=}N$ and $a_{12}=a_{21}\overset{\triangle}{=}a$.
We have

\begin{equation}
\xi\left(P,N,a\right)=\frac{1}{2}\log\left(1+\frac{aP}{N}\right)-u.c.e\left\{ \left[\frac{1}{2}\log\left(\frac{P\left(1+a\right)+N}{2N+P\left(1-\sqrt{a}\right)^{2}}\right)\right]^{+}\right\} ,\label{eq:Gap for symmetric model}
\end{equation}
 and the condition on noise variance is reduced to

\[
N\geq\left(\sqrt{a}-1\right)P,
\]
 where $a\geq1$ (strong interference). Now, we investigate the second
term for obtaining its minimum value. We can see that gap is an increasing
function of $a$ for $1\leq a\leq\left(\frac{P+N}{P}\right)^{2}$.
Therefore, its maximum value occurs at $a=\left(\frac{P+N}{P}\right)^{2}$.
Thus, to obtain the gap, we evaluate it for $a=\left(\frac{P+N}{P}\right)^{2}$.
We have

\begin{equation}
\xi\left(P,N,\left(\frac{P+N}{P}\right)^{2}\right)=\frac{1}{2}\log\left(1+\frac{\left(P+N\right)^{2}}{NP}\right)-u.c.e\left\{ \left[\frac{1}{2}\log\left(\frac{2P^{2}+3PN+N^{2}}{2PN+N^{2}}\right)\right]^{+}\right\} .\label{eq:Gap for symmetric model a=00003D00003D1}
\end{equation}
 Let us define $x$ as $x\overset{\triangle}{=}\frac{P}{N}.$ Thus,

\begin{equation}
\xi\left(P,N,\left(\frac{P+N}{P}\right)^{2}\right)=\frac{1}{2}\log\left(1+\frac{\left(x+1\right)^{2}}{x}\right)-u.c.e\left\{ \left[\frac{1}{2}\log\left(\frac{2x^{2}+3x+1}{2x+1}\right)\right]^{+}\right\} \overset{\Delta}{=}\tilde{\xi}\left(x\right).\label{eq:Gap between inner bound and outer bound}
\end{equation}
 As we can see in Fig. \ref{fig:Gap-is-given}, $\tilde{\xi}\left(x\right)$
is a decreasing function of $x$. Thus, it is maximized at $x\rightarrow0$.
Unfortunately, the gap tends to infinity as $x\rightarrow0$. But,
as we can see in Table \ref{tab:The-gap-between}, the gap is smaller
than 0.67 bit for $x\geq1$ and tends to zero as $x\rightarrow\infty$.

\begin{table}
\centering{}\caption{{\small \label{tab:The-gap-between}The gap between the outer bound
and the achievable rate.}}

\centering{}\centering{}%
\begin{tabular}{|c|c|c|c|c|c|}
\hline 
SNR  & .1  & .5  & 1  & 10  & 20 \tabularnewline
\hline 
Gap (bits)  & 1.79  & 0.938  & 0.661  & 0.1257  & 0.0673 \tabularnewline
\hline 
\end{tabular}
\end{table}

Note that, we can theorem (\ref{thm:Imbalanced-SNRs:-Suppose}) and
theorem (\ref{thm:If--and-1}) to obtain the following achievable
regions for other conditions over both noise variances. 
\begin{cor}
\label{thm:If--and-1} If $N_{1}\leq\sqrt{a_{12}P_{2}P_{1}}-a_{12}P_{2}$
and $N_{2}\geq\sqrt{a_{21}P_{2}P_{1}}-\min\left(a_{21}P_{1},P_{2}\right)$
for $P_{1}\neq a_{12}P_{2}$, $a_{21}P_{1}\neq P_{2}$ and $P_{1},P_{2}\geq1$,
then the following region is achievable for ASD-GIC:

\begin{equation}
R_{1}+R_{2}\leq\min\left(\frac{1}{2}\log\left(1+\frac{a_{12}P_{2}}{N_{1}}\right),u.c.e\left\{ \left[\frac{1}{2}\log\left(\frac{P_{2}+a_{21}P_{1}+N_{2}}{2N_{2}+\left(\sqrt{P_{2}}-\sqrt{a_{21}}P_{1}\right)^{2}}\right)\right]^{+}\right\} \right),\label{eq:Achievable rate for other condition over nosie 1}
\end{equation}
 where the upper convex envelope is with respect to $P_{1}$ and $P_{2}$.\end{cor}
\begin{IEEEproof}
By using (\ref{eq:achievable rate for decoder 1}) and (\ref{eq:Achievable region for decoder 2}),
the proof is straightforward.\end{IEEEproof}
\begin{cor}
\label{thm:If--and-2} If $N_{1}\geq\sqrt{a_{12}P_{2}P_{1}}-\min\left(a_{12}P_{2},P_{1}\right)$
and $N_{2}\leq\sqrt{a_{21}P_{2}P_{1}}-a_{21}P_{1}$ for $P_{1}\neq a_{12}P_{2}$,
$a_{21}P_{1}\neq P_{2}$ and $P_{1},P_{2}\geq1$, then the following
region is achievable for ASD-GIC:

\begin{equation}
R_{1}+R_{2}\leq\min\left(u.c.e\left\{ \left[\frac{1}{2}\log\left(\frac{P_{1}+a_{12}P_{2}+N_{1}}{2N_{1}+\left(\sqrt{P_{1}}-\sqrt{a_{12}P_{2}}\right)^{2}}\right)\right]^{+}\right\} ,\frac{1}{2}\log\left(1+\frac{a_{21}P_{1}}{N_{2}}\right)\right),\label{eq:Achievable rate for other condition over nosie 1-1}
\end{equation}
 where the upper convex envelope is with respect to $P_{1}$ and $P_{2}$.\end{cor}
\begin{IEEEproof}
Using (\ref{eq:achievable rate for decoder 2}) and (\ref{eq:Achievable region for Decoder 1}),
the proof is straightforward. 
\end{IEEEproof}

\section{\label{sec:Conclusion}Conclusion}

In this paper, an additive state-dependent Gaussian interference channel
(ASD-GIC) is considered. We assume that the state power goes to infinity.
We establish four achievable rate regions by using lattice coding
scheme. Dependent on noise variances, we reach to capacity or to a
constant gap in the symmetric model.

\appendix
{}

\section*{\label{sec:Appendix1}Appendix 1}
\begin{prop}
For both MACs in ASD-GIC, in the limit of high SNR, where $SNR_{1}=\frac{P_{1}}{N_{1}}\gg1$
and $SNR_{2}=\frac{P_{2}}{N_{2}}\gg1$, the achievable sum-rate can
be upper bounded by

\begin{equation}
R_{1}+R_{2}\leq\left[h(S_{1}+S_{2})-h(S_{1})-h(S_{2})+\frac{1}{2}\log\left(\frac{2\pi eP_{1}P_{2}}{N_{1}}\right)\right]^{+}.\label{eq:Sum rate by random binning}
\end{equation}
 \end{prop}
\begin{IEEEproof}
We consider MAC 1. As for Costa dirty paper coding, we choose auxiliary
random variables $U_{1}$ and $U_{2}$ as

\begin{eqnarray*}
U_{1} & = & X_{1}+\alpha_{1}S_{1},\\
U_{2} & = & X_{2}+\alpha_{2}S_{2},
\end{eqnarray*}
 where $\alpha_{1}=\frac{P_{1}}{P_{1}+N_{1}}$ and $\alpha_{2}=\frac{a_{12}P_{2}}{a_{12}P_{2}+N_{1}}$.
In the limit of high SNR, where $SNR_{1}=\frac{P_{1}}{N_{1}}\gg1$
and $SNR_{2}=\frac{P_{2}}{N_{2}}\gg1$, we have $\alpha_{1}\thickapprox1,\alpha_{2}\thickapprox1$.
Thus, $U_{1}=X_{1}+S_{1},$ and $U_{2}=X_{2}+S_{2}$. Now, by using
these auxiliary random variables in sum-rate provided in (\ref{eq:Sum rate In Jafar region}),
we have

\begin{eqnarray*}
R_{1}+R_{2} & = & \left[I\left(U_{1},U_{2};Y_{1}\right)-I(U_{1};S_{1})-I(U_{2};S_{2})\right]^{+},\\
 & = & \left[h\left(Y_{1}\right)-h(Y_{1}|U_{1},U_{2})-h(U_{1})+h(X_{1})-h(U_{2})+h(X_{2})\right]^{+},\\
 & \leq & \left[h\left(Y_{1}\right)-h(Z_{1})-h(S_{1})+h(X_{1})-h(S_{2})+h(X_{2})\right]^{+},\\
 & = & \left[h\left(Y_{1}\right)-h(S_{1})-h(S_{2})+\Gamma\right]^{+},\\
 & \leq & \left[h\left(S_{1}+S_{2}\right)-h(S_{1})-h(S_{2})+\Gamma\right]^{+},
\end{eqnarray*}
 where $[x]^{+}=\max\left\{ x,0\right\} $ and $\Gamma=\frac{1}{2}\log\left(\frac{2\pi eP_{1}P_{2}}{N_{1}}\right)$.
For MAC 2, we can obtain similar result. 
\end{IEEEproof}
Now, by evaluating the upper bound in (\ref{eq:Sum rate by random binning})
for $Q_{i}\rightarrow\infty$, we get

\[
\underset{Q_{i}\rightarrow\infty}{\lim}\left[h\left(S_{1}+S_{2}\right)-h(S_{1})-h(S_{2})+\Gamma\right]^{+}=\underset{Q_{i}\rightarrow\infty}{\lim}\left[\frac{1}{2}\log\left(\frac{Q_{1}+Q_{2}}{Q_{1}Q_{2}}\right)+\Gamma\right]^{+}\rightarrow0.
\]
 Thus, for $Q_{i}\rightarrow\infty$, we cannot reach any positive
rate by random binning scheme.

\section*{\label{sec:Appendix-2}Appendix 2}

Here, we obtain the following corner point:

\[
\left(R_{1},R_{2}\right)=\left(\left[\frac{1}{2}\log\left(\frac{P_{1}+a_{12}P_{2}+N_{1}}{2N_{1}+\left(\sqrt{P_{1}}-\sqrt{a_{12}P_{2}}\right)^{2}}\right)\right]^{+},0\right).
\]
 We assume that $\Lambda_{1}$ and $\Lambda_{2}$ are two lattices,
which are good for quantization, with second moments $P_{1}$ and
$P_{2}$, respectively.

First, we consider $P_{1}\leq\frac{\left(a_{12}P_{2}+N_{1}\right)^{2}}{a_{12}P_{2}}$
. For this case, we assume that $\frac{\alpha_{1}}{\alpha_{2}}\sqrt{a_{12}}\Lambda_{2}=\Lambda_{1}$.
The encoders send 
\begin{eqnarray*}
\boldsymbol{X}_{1} & = & \left[\boldsymbol{V}_{1}-\alpha_{1}\boldsymbol{S}_{1}+\boldsymbol{D}_{1}\right]\textrm{ mod }\Lambda_{1},\\
\boldsymbol{X}_{2} & = & \left[-\alpha_{2}\boldsymbol{S}_{2}+\boldsymbol{D}_{2}\right]\textrm{ mod }\Lambda_{2}.
\end{eqnarray*}
 At the receiver of user 1, based on the channel output, given by
\[
\boldsymbol{Y}_{1}=\boldsymbol{X}_{1}+\sqrt{a_{12}}\boldsymbol{X}_{2}+\boldsymbol{S}_{1}+\sqrt{a_{12}}\boldsymbol{S}_{2}+\boldsymbol{Z}_{1},
\]
 the following operation is performed: 
\begin{eqnarray}
\boldsymbol{Y}_{d1} & = & \left[\alpha_{1}\boldsymbol{Y}_{1}-\sqrt{a_{12}}\frac{\alpha_{1}}{\alpha_{2}}\boldsymbol{D}_{2}-\boldsymbol{D}_{1}\right]\textrm{ mod }\Lambda_{1},\nonumber \\
 & = & \left[\alpha_{1}\left(\boldsymbol{X}_{1}+\sqrt{a_{12}}\left[-\alpha_{2}\boldsymbol{S}_{2}+\boldsymbol{D}_{2}\right]\textrm{ mod }\Lambda_{2}+\boldsymbol{S}_{1}+\sqrt{a_{12}}\boldsymbol{S}_{2}+\boldsymbol{Z}_{1}\right)-\sqrt{a_{12}}\frac{\alpha_{1}}{\alpha_{2}}\boldsymbol{D}_{2}-\boldsymbol{D}_{1}\right]\textrm{ mod }\Lambda_{1},\nonumber \\
 & = & \left[\boldsymbol{V}_{1}+\alpha_{1}\left(\boldsymbol{X}_{1}+\boldsymbol{Z}_{1}\right)-\left(\boldsymbol{V}_{1}-\alpha_{1}\boldsymbol{S}_{1}+\boldsymbol{D}_{1}\right)-\sqrt{a_{12}}\left(1-\alpha_{2}\right)\frac{\alpha_{1}}{\alpha_{2}}\left(-\alpha_{2}\boldsymbol{S}_{2}+\boldsymbol{D}_{2}\right)\right.\nonumber \\
 &  & \left.-\sqrt{a_{12}}\alpha_{1}\mathcal{Q}_{\Lambda_{2}}\left(-\alpha_{2}\boldsymbol{S}_{2}+\boldsymbol{D}_{2}\right)\right]\textrm{ mod }\Lambda_{1},\nonumber \\
 & = & \left[\boldsymbol{V}_{1}+\left(\alpha_{1}-1\right)\boldsymbol{X}_{1}-\sqrt{a_{12}}\left(1-\alpha_{2}\right)\frac{\alpha_{1}}{\alpha_{2}}\boldsymbol{X}_{2}+\alpha_{1}\boldsymbol{Z}_{1}-\sqrt{a_{12}}\frac{\alpha_{1}}{\alpha_{2}}\mathcal{Q}_{\Lambda_{2}}\left(-\alpha_{2}\boldsymbol{S}_{2}+\boldsymbol{D}_{2}\right)\right]\textrm{ mod }\Lambda_{1},\label{eq:Distributive law for R_1 (Imbalanced ..)}\\
 & = & \left[\boldsymbol{V}_{1}+\left(\alpha_{1}-1\right)\boldsymbol{X}_{1}-\sqrt{a_{12}}\left(1-\alpha_{2}\right)\frac{\alpha_{1}}{\alpha_{2}}\boldsymbol{X}_{2}+\alpha_{1}\boldsymbol{Z}_{1}\right]\textrm{ mod }\Lambda_{1},\label{eq:Lattice Alignment for R_1 (Imbalanced ..)}\\
 & = & \left[\boldsymbol{V}_{1}+\boldsymbol{Z}_{eff}\right]\textrm{ mod }\Lambda_{1},\nonumber 
\end{eqnarray}
 where 
\[
\boldsymbol{Z}_{eff}=\left[\left(\alpha_{1}-1\right)\boldsymbol{X}_{1}-\sqrt{a_{12}}\left(1-\alpha_{2}\right)\frac{\alpha_{1}}{\alpha_{2}}\boldsymbol{X}_{2}+\alpha_{1}\boldsymbol{Z}_{1}\right]\textrm{ mod }\Lambda_{1}.
\]
 (\ref{eq:Distributive law for R_1 (Imbalanced ..)}) is based on
distributive law and (\ref{eq:Lattice Alignment for R_1 (Imbalanced ..)})
follows from $\frac{\alpha_{1}}{\alpha_{2}}\sqrt{a_{12}}\Lambda_{2}=\Lambda_{1}$,
we have that $\sqrt{a_{12}}\frac{\alpha_{1}}{\alpha_{2}}\mathcal{Q}_{\Lambda_{2}}\left(\alpha_{2}\boldsymbol{S}_{2}+\boldsymbol{D}_{2}\right)\in\Lambda_{1}$,
i.e., the interference signal is aligned with $\Lambda_{1}$. To calculate
the rate $R_{1}$, it is assumed that $\boldsymbol{V}_{1}\sim\mbox{ Unif}\left(\mathcal{V}_{1}\right)$.
We have 
\begin{eqnarray}
R_{1} & = & \frac{1}{n}I\left(\boldsymbol{V}_{1};\boldsymbol{Y}_{d1}\right),\nonumber \\
 & = & \frac{1}{n}\left\{ h(\boldsymbol{Y}_{d1})-h(\boldsymbol{Y}_{d1}|\boldsymbol{V}_{1})\right\} \nonumber \\
 & = & \frac{1}{2}\log\left(\frac{P_{1}}{G\left(\Lambda_{1}\right)}\right)-\frac{1}{n}h\left(\left[\left(\alpha_{1}-1\right)\boldsymbol{X}_{1}-\sqrt{a_{12}}\left(1-\alpha_{2}\right)\frac{\alpha_{1}}{\alpha_{2}}\boldsymbol{X}_{2}+\alpha_{1}\boldsymbol{Z}_{1}\right]\textrm{ mod }\Lambda_{1}\right),\label{eq:Crypto lemma for R_1(Imbalanced)}\\
 & \geq & \frac{1}{2}\log\left(\frac{P_{1}}{\left(\alpha_{1}-1\right)^{2}P_{1}+\left(\left(1-\alpha_{2}\right)\frac{\alpha_{1}}{\alpha_{2}}\right)^{2}a_{12}P_{2}+\alpha_{1}^{2}N_{1}}\right)-\frac{1}{2}\log\left(2\pi eG\left(\Lambda_{1}\right)\right),\label{eq:Gaussian Entropy for R_1 (Imbalanced)}
\end{eqnarray}
 where (\ref{eq:Crypto lemma for R_1(Imbalanced)}) follows from the
fact that $\sqrt{a_{12}}\boldsymbol{V}_{2}$ is uniform over $\sqrt{a_{12}}\mathcal{V}_{2}$,
thus so $\boldsymbol{Y}_{d1}$ is uniform over $\sqrt{a_{12}}\mathcal{V}_{2}$
(crypto lemma). (\ref{eq:Gaussian Entropy for R_1 (Imbalanced)})
follows from the fact that modulo operation reduces the second moment
and Gaussian distribution maximizes differential entropy for a fixed
second moment. Now, by considering $\left(\frac{\alpha_{1}}{\alpha_{2}}\right)^{2}a_{12}P_{2}=P_{1}$,
we find the optimal $\alpha$ when the lattice dimension goes to infinity
such that the MSE of the effective noise $\boldsymbol{Z}_{{\rm eff}}$
is minimized. Hence, 
\[
\alpha_{{\rm 1,MMSE}}=\frac{\sqrt{P_{1}}\left(\sqrt{P_{1}}+\sqrt{a_{12}P_{2}}\right)}{P_{1}+a_{12}P_{2}+N_{1}}.
\]
 With this $\alpha$, we get that the following achievable rate:

\begin{equation}
R_{1}\leq\left[\frac{1}{2}\log\left(\frac{P_{1}+a_{12}P_{2}+N_{1}}{2N_{1}+\left(\sqrt{P_{1}}-\sqrt{a_{12}P_{2}}\right)^{2}}\right)\right]^{+}.\label{eq:Case 1, Achievable rate 1-1}
\end{equation}
 Thus, if $P_{1}\leq\frac{\left(a_{12}P_{2}+N_{1}\right)^{2}}{a_{12}P_{2}}$,
we can achieve the following corner point:

\begin{equation}
\left(R_{1},R_{2}\right)=\left(u.c.e\left\{ \left[\frac{1}{2}\log\left(\frac{P_{1}+a_{12}P_{2}+N_{1}}{2N_{1}+\left(\sqrt{P_{1}}-\sqrt{a_{12}P_{2}}\right)^{2}}\right)\right]^{+}\right\} ,0\right)\label{eq:Corner Point 1 for R_1 (Imbalanced ..)}
\end{equation}
 Now, we consider $a_{12}P_{2}\leq\frac{\left(P_{1}+N_{1}\right)^{2}}{P_{1}}$
. For this case, we assume that $\Lambda_{3}=\frac{\alpha_{2}}{\alpha_{1}}\Lambda_{1}$,
where $\Lambda_{3}=\sqrt{a_{12}}\Lambda_{2}$. The encoders send 
\begin{eqnarray*}
\boldsymbol{X}_{1} & = & \left[\boldsymbol{V}_{1}-\alpha_{1}\boldsymbol{S}_{1}+\boldsymbol{D}_{1}\right]\textrm{ mod }\Lambda_{1},\\
\boldsymbol{X}_{2} & = & \left[-\alpha_{2}\boldsymbol{S}_{2}+\boldsymbol{D}_{2}\right]\textrm{ mod }\Lambda_{2}.
\end{eqnarray*}
 At the receiver of user 1, based on the channel output given by 
\[
\boldsymbol{Y}_{1}=\boldsymbol{X}_{1}+\sqrt{a_{12}}\boldsymbol{X}_{2}+\boldsymbol{S}_{1}+\sqrt{a_{12}}\boldsymbol{S}_{2}+\boldsymbol{Z}_{1},
\]
 the following operation is performed: 
\begin{eqnarray}
\boldsymbol{Y}_{d1} & = & \left[\alpha_{2}\boldsymbol{Y}_{1}-\sqrt{a_{12}}\boldsymbol{D}_{2}-\frac{\alpha_{2}}{\alpha_{1}}\boldsymbol{D}_{1}\right]\textrm{ mod }\Lambda_{3},\nonumber \\
 & = & \left[\alpha_{2}\left(\left[\boldsymbol{V}_{1}-\alpha_{1}\boldsymbol{S}_{1}+\boldsymbol{D}_{1}\right]\textrm{ mod }\Lambda_{1}+\sqrt{a_{12}}\boldsymbol{X}_{2}+\boldsymbol{S}_{1}+\sqrt{a_{12}}\boldsymbol{S}_{2}+\boldsymbol{Z}_{1}\right)-\sqrt{a_{12}}\boldsymbol{D}_{2}-\frac{\alpha_{2}}{\alpha_{1}}\boldsymbol{D}_{1}\right]\textrm{ mod }\Lambda_{3},\nonumber \\
 & = & \left[\frac{\alpha_{2}}{\alpha_{1}}\boldsymbol{V}_{1}+\alpha_{2}\left(\sqrt{a_{12}}\boldsymbol{X}_{2}+\boldsymbol{Z}_{1}\right)-\sqrt{a_{12}}\left(-\alpha_{2}\boldsymbol{S}_{2}+\boldsymbol{D}_{2}\right)-\left(1-\alpha_{1}\right)\frac{\alpha_{2}}{\alpha_{1}}\left(\boldsymbol{V}_{1}-\alpha_{1}\boldsymbol{S}_{1}+\boldsymbol{D}_{1}\right)\right.\nonumber \\
 &  & \left.-\alpha_{2}\mathcal{Q}_{\Lambda_{1}}\left(\boldsymbol{V}_{1}-\alpha_{1}\boldsymbol{S}_{1}+\boldsymbol{D}_{1}\right)\right]\textrm{ mod }\Lambda_{3},\nonumber \\
 & = & \left[\frac{\alpha_{2}}{\alpha_{1}}\boldsymbol{V}_{1}+\sqrt{a_{12}}\left(\alpha_{2}-1\right)\boldsymbol{X}_{2}-\left(1-\alpha_{1}\right)\frac{\alpha_{2}}{\alpha_{1}}\boldsymbol{X}_{1}+\alpha_{2}\boldsymbol{Z}_{1}-\frac{\alpha_{2}}{\alpha_{1}}\mathcal{Q}_{\Lambda_{1}}\left(\boldsymbol{V}_{1}-\alpha_{1}\boldsymbol{S}_{1}+\boldsymbol{D}_{1}\right)\right]\textrm{ mod }\Lambda_{3},\label{eq:Distributive law for R_1 (Imbalanced ..)-1}\\
 & = & \left[\frac{\alpha_{2}}{\alpha_{1}}\boldsymbol{V}_{1}+\sqrt{a_{12}}\left(\alpha_{2}-1\right)\boldsymbol{X}_{2}-\left(1-\alpha_{1}\right)\frac{\alpha_{2}}{\alpha_{1}}\boldsymbol{X}_{1}+\alpha_{2}\boldsymbol{Z}_{1}\right]\textrm{ mod }\Lambda_{3},\label{eq:Lattice Alignment for R_1 (Imbalanced ..)-1}\\
 & = & \left[\frac{\alpha_{2}}{\alpha_{1}}\boldsymbol{V}_{1}+\boldsymbol{Z}_{eff}\right]\textrm{ mod }\Lambda_{3},\nonumber 
\end{eqnarray}
 where 
\[
\boldsymbol{Z}_{eff}=\left[\sqrt{a_{12}}\left(\alpha_{2}-1\right)\boldsymbol{X}_{2}-\left(1-\alpha_{1}\right)\frac{\alpha_{2}}{\alpha_{1}}\boldsymbol{X}_{1}+\alpha_{2}\boldsymbol{Z}_{1}\right]\textrm{ mod }\Lambda_{3}.
\]
 (\ref{eq:Distributive law for R_1 (Imbalanced ..)-1}) is based on
distributive law and (\ref{eq:Lattice Alignment for R_1 (Imbalanced ..)-1})
follows from $\frac{\alpha_{2}}{\alpha_{1}}\Lambda_{1}=\Lambda_{3}$,
we have that $\frac{\alpha_{2}}{\alpha_{1}}\mathcal{Q}_{\Lambda_{1}}\left(\alpha_{2}\boldsymbol{S}_{2}+\boldsymbol{D}_{2}\right)\in\Lambda_{3}$,
i.e., the interference signal is aligned with $\Lambda_{3}$. Hence,
the element disappears after the modulo operation. To calculate rate
$R_{1}$, it is assumed that $\boldsymbol{V}_{1}\sim\mbox{ Unif}\left(\mathcal{V}_{1}\right)$.
We have 
\begin{eqnarray}
R_{1} & = & \frac{1}{n}I\left(\boldsymbol{V}_{1};\boldsymbol{Y}_{d1}\right),\nonumber \\
 & = & \frac{1}{n}\left\{ h(\boldsymbol{Y}_{d1})-h(\boldsymbol{Y}_{d1}|\boldsymbol{V}_{1})\right\} \nonumber \\
 & = & \frac{1}{2}\log\left(\frac{a_{12}P_{2}}{G\left(\Lambda_{3}\right)}\right)-\frac{1}{n}h\left(\left[\sqrt{a_{12}}\left(\alpha_{2}-1\right)\boldsymbol{X}_{2}-\left(1-\alpha_{1}\right)\frac{\alpha_{2}}{\alpha_{1}}\boldsymbol{X}_{1}+\alpha_{2}\boldsymbol{Z}_{1}\right]\textrm{ mod \ensuremath{\Lambda_{3}}}\right),\label{eq:Crypto lemma for R_1(Imbalanced)-1}\\
 & \geq & \frac{1}{2}\log\left(\frac{a_{12}P_{2}}{\left(\alpha_{2}-1\right)^{2}a_{12}P_{1}+\left(\left(1-\alpha_{1}\right)\frac{\alpha_{2}}{\alpha_{1}}\right)^{2}P_{1}+\alpha_{2}^{2}N_{1}}\right)-\frac{1}{2}\log\left(2\pi eG\left(\Lambda_{3}\right)\right),\label{eq:Gaussian Entropy for R_1 (Imbalanced)-1}
\end{eqnarray}
 Since $\frac{\alpha_{2}}{\alpha_{1}}\boldsymbol{V}_{1}$ is uniform
over $\frac{\alpha_{2}}{\alpha_{1}}\mathcal{V}_{2}$, $\boldsymbol{Y}_{d1}$
is also uniform over $\frac{\alpha_{2}}{\alpha_{1}}\mathcal{V}_{2}$
(crypto lemma), thus (\ref{eq:Crypto lemma for R_1(Imbalanced)-1})
is correct. (\ref{eq:Gaussian Entropy for R_1 (Imbalanced)-1}) follows
from the fact that modulo operation reduces the second moment and
Gaussian distribution maximizes differential entropy for a fixed second
moment. Now, by considering $\left(\frac{\alpha_{2}}{\alpha_{1}}\right)^{2}P_{1}=a_{12}P_{2}$,
and the MMSE value of $\alpha$, which minimizes the MSE of the effective
noise, $\boldsymbol{Z}_{{\rm eff}}$, 
\[
\alpha_{{\rm 2,MMSE}}=\frac{\sqrt{a_{12}P_{2}}\left(\sqrt{P_{1}}+\sqrt{a_{12}P_{2}}\right)}{P_{1}+a_{12}P_{2}+N_{1}}.
\]
 we get the following achievable rate:

\begin{equation}
R_{1}\leq\left[\frac{1}{2}\log\left(\frac{P_{1}+a_{12}P_{2}+N_{1}}{2N_{1}+\left(\sqrt{P_{1}}-\sqrt{a_{12}P_{2}}\right)^{2}}\right)\right]^{+}.\label{eq:Case 1, Achievable rate 1-1-1}
\end{equation}
 Thus, if $a_{12}P_{2}\leq\frac{\left(P_{1}+N_{1}\right)^{2}}{P_{1}}$
, then we can achieve the following corner point:

\begin{equation}
\left(R_{1},R_{2}\right)=\left(u.c.e\left\{ \left[\frac{1}{2}\log\left(\frac{P_{1}+a_{12}P_{2}+N_{1}}{2N_{1}+\left(\sqrt{P_{1}}-\sqrt{a_{12}P_{2}}\right)^{2}}\right)\right]^{+}\right\} ,0\right)\label{eq:Corner Point 1 for R_1 (Imbalanced ..)-1}
\end{equation}
 Now, by combining (\ref{eq:Corner Point 1 for R_1 (Imbalanced ..)})
and (\ref{eq:Corner Point 1 for R_1 (Imbalanced ..)-1}), we get the
following corner point

\[
\left(R_{1},R_{2}\right)=\left(u.c.e\left\{ \left[\frac{1}{2}\log\left(\frac{P_{1}+a_{12}P_{2}+N_{1}}{2N_{1}+\left(\sqrt{P_{1}}-\sqrt{a_{12}P_{2}}\right)^{2}}\right)\right]^{+}\right\} ,0\right),
\]
 if

\[
N_{1}\geq\sqrt{a_{12}P_{1}P_{2}}-\min\left(a_{12}P_{2},P_{1}\right).
\]

 \bibliographystyle{IEEEtran}
\bibliography{IEEEabrv,ReferencesGhasemiDec2012}

\begin{thebibliography}{10}
\providecommand{\url}[1]{#1}
\csname url@samestyle\endcsname
\providecommand{\newblock}{\relax}
\providecommand{\bibinfo}[2]{#2}
\providecommand{\BIBentrySTDinterwordspacing}{\spaceskip=0pt\relax}
\providecommand{\BIBentryALTinterwordstretchfactor}{4}
\providecommand{\BIBentryALTinterwordspacing}{\spaceskip=\fontdimen2\font plus
\BIBentryALTinterwordstretchfactor\fontdimen3\font minus
  \fontdimen4\font\relax}
\providecommand{\BIBforeignlanguage}[2]{{%
\expandafter\ifx\csname l@#1\endcsname\relax
\typeout{** WARNING: IEEEtran.bst: No hyphenation pattern has been}%
\typeout{** loaded for the language `#1'. Using the pattern for}%
\typeout{** the default language instead.}%
\else
\language=\csname l@#1\endcsname
\fi
#2}}
\providecommand{\BIBdecl}{\relax}
\BIBdecl

\bibitem{Zhang_Allerton_2011}
L.~Zhang, T.~Liu, and S.~Cui, ``Symmetric {G}aussian interference channel with
  state information,'' in \emph{Proc. 49th Annual Allerton Conference on
  Communication, Control, and Computing (Allerton)}, Sep. 2011, pp. 832--838.

\bibitem{Zhang_IT_11}
L.~Zhang, J.~Jiang, and S.~Cui, ``Interference channel with state
  information,'' \emph{{IEEE} Trans. Inf. Theory}, submitted for publication,
  April. 2011.

\bibitem{Philosof_11}
T.~Philosof, R.~Zamir, U.~Erez, and A.~J. Khisti, ``Lattice strategies for the
  dirty multiple access channel,'' \emph{{IEEE} Trans. Inf. Theory}, vol.~57,
  no.~8, pp. 5006--5035, Aug. 2011.

\bibitem{Shannon_61}
C.~E. Shannon, ``Two way communication channels,'' in \emph{Proc. 4th Berkeley
  Symp. on Mathematical Statistics and Probability}, Berkeley, CA, 1961, pp.
  611--644.

\bibitem{Sato_IT_81}
H.~Sato, ``The capacity of the {G}aussian interference channel under strong
  interference,'' \emph{{IEEE} Trans. Inf. Theory}, vol.~27, no.~6, pp.
  786--788, Nov. 1981.

\bibitem{Carleial_IT_75}
A.~B. Carleial, ``A case where interference does not reduce capacity,''
  \emph{{IEEE} Trans. Inf. Theory}, vol.~21, no.~5, pp. 569--570, Sep. 1975.

\bibitem{Sason_IT_2004}
I.~Sason, ``On achievable rate regions for the {G}aussian interference
  channel,'' \emph{{IEEE} Trans. Inf. Theory}, vol.~53, no.~12, pp. 1345--1356,
  2004.

\bibitem{Carleial_IT_78}
A.~B. Carleial, ``Interference channels,'' \emph{{IEEE} Trans. Inf. Theory},
  vol.~24, no.~1, pp. 60--70, Jan. 1978.

\bibitem{Han_IT_81}
T.~S. Han and K.~Kobayashi, ``A new achievable rate region for the interference
  channel,'' \emph{{IEEE} Trans. Inf. Theory}, vol.~27, no.~1, pp. 49--60, Jan.
  1981.

\bibitem{Etkin_IT_2008}
R.~H. Etkin, D.~N.~C. Tse, and H.~Wang, ``Gaussian interference channel
  capacity to within one bit,'' \emph{{IEEE} Trans. Inf. Theory}, vol.~54,
  no.~12, pp. 5534--5562, Dec. 2008.

\bibitem{Maric_IT_2007}
I.~Maric, R.~D. Yates, and G.~Kramer, ``Capacity of interference channels with
  partial transmitter cooperation,'' \emph{{IEEE} Trans. Inf. Theory}, vol.~53,
  no.~10, pp. 3536--3548, Oct. 2007.

\bibitem{Cao_07_ISIT}
Y.~Cao and B.~Chen, ``An achievable rate region for interference channels with
  conferencing,'' in \emph{Proc. IEEE ISIT}, Nice, France, Jun. 2007, pp.
  1251--1255.

\bibitem{Prabhakaran_IT_Cooperation_2011}
V.~Prabhakaran and P.~Viswanath, ``Interference channels with source
  cooperation,'' \emph{{IEEE} Trans. Inf. Theory}, vol.~57, no.~1, pp.
  156--186, Jan. 2011.

\bibitem{Changho_IT_2011}
S.~Changho and D.~N.~C. Tse, ``Feedback capacity of the {G}aussian interference
  channel to within 2 bits,'' \emph{{IEEE} Trans. Inf. Theory}, vol.~57, no.~5,
  pp. 2667--2685, May 2011.

\bibitem{Tian_IT_11}
Y.~Tian and A.~Yener, ``The {G}aussian interference relay channel: improved
  achievable rates and sum rate upperbounds using a potent relay,''
  \emph{{IEEE} Trans. Inf. Theory}, vol.~57, no.~5, pp. 2865--2879, May 2011.

\bibitem{Zhang_WCNC_2011}
L.~Zhang, J.~Jiang, and S.~Cui, ``Gaussian interference channel with state
  information,'' in \emph{Proc. 2011 IEEE Wireless Communications and
  Networking Conference (WCNC)}, Mar. 2011, pp. 1960--1965.

\bibitem{Zamir_09_ITA}
R.~Zamir, ``Lattices are everywhere,'' in \emph{Proceedings of the 4th Annual
  Workshop on Information Theory and its Applications (ITA 2009)}, San Diego,
  CA, Feb. 2009, pp. 392 -- 421.

\bibitem{Erez_IT_04}
U.~Erez and R.~Zamir, ``Achieving 1/2 log(1+{SNR}) on the {AWGN} channel with
  lattice encoding and decoding,'' \emph{{IEEE} Trans. Inf. Theory}, vol.~50,
  no.~22, pp. 2293--2314, Oct. 2004.

\bibitem{Ershi_IT_05}
U.~Erez, S.Shamai, and R.~Zamir, ``Capacity and lattice strategies for
  canceling known interference,'' \emph{{IEEE} Trans. Inf. Theory}, vol.~51,
  no.~14, pp. 3820--3833, Nov. 2005.

\bibitem{Costa_83}
M.~Costa, ``Writing on dirty paper,'' \emph{{IEEE} Trans. Inf. Theory},
  vol.~29, no.~3, pp. 439--441, May 1983.

\bibitem{Wang_IT_2012}
I.-H. Wang, ``Approximate capacity of the dirty multiple-access channel with
  partial state information at the encoders,'' \emph{{IEEE} Trans. Inf.
  Theory}, vol.~58, no.~5, pp. 2781--2787, May 2012.

\bibitem{Ghasemi_12_PIMRC}
S.~Ghasemi-Goojani and H.~Behroozi, ``On the sum-capacity and lattice-based
  transmission strategies for state-dependent {G}aussian interference
  channel,'' in \emph{Proc. 23th IEEE Int. Symp. on Personal, Indoor and Mobile
  Radio Communications (PIMRC)}, Sydney, Australia, Sep. 2012.

\bibitem{Cover_book_2ndEdition}
T.~M. Cover and J.~A. Thomas, \emph{Elements of Information Theory}.\hskip 1em
  plus 0.5em minus 0.4em\relax New York: 2nd Edition, John Wiley \& Sons, 2006.

\bibitem{Jafar_IT_06}
S.~A. Jafar, ``Capacity with causal and noncausal side information-a unified
  view,'' \emph{{IEEE} Trans. Inf. Theory}, vol.~52, no.~12, pp. 5468--5475,
  Dec. 2006.

\bibitem{Nazar_IT_11}
B.~Nazer and M.~Gastpar, ``{C}ompute-and-forward: {H}arnessing interference
  through structured codes,'' \emph{{IEEE} Trans. Inf. Theory}, vol.~57,
  no.~10, pp. 6463--6486, Oct. 2011.

\bibitem{conway_book}
J.~H. Conway and N.~J.~A. Sloane, \emph{Sphere Packings, Lattices and
  Groups}.\hskip 1em plus 0.5em minus 0.4em\relax New York: Springer-Verlag,
  1992.

\bibitem{Zamir_IT_96}
R.~Zamir and M.~Feder, ``On lattice quantization noise,'' \emph{{IEEE} Trans.
  Inf. Theory}, vol.~42, no.~4, pp. 1152--1159, Jul. 1996.

\bibitem{Erez_IT_05}
U.~Erez, S.~Litsyn, and R.~Zamir, ``Lattices which are good for (almost)
  everything,'' \emph{{IEEE} Trans. Inf. Theory}, vol.~51, no.~16, pp.
  3401--3416, Oct. 2005.

\bibitem{Poltyrev_IT_94}
G.~Poltyrev, ``On coding without restrictions for the {AWGN} channel,''
  \emph{{IEEE} Trans. Inf. Theory}, vol.~40, no.~9, pp. 409 -- 417, Mar. 1994.

\bibitem{Forney_Allerton_2003}
G.~D. Forney, ``On the role of {MMSE} estimation in approaching the information
  theoretic limits of linear {G}aussian channels: Shannon meets {W}iener,'' in
  \emph{Proc. 41st Ann. Allerton Conf.}

\end{thebibliography}

\end{document}